\documentclass[pdflatex,sn-mathphys-num]{sn-jnl}
\usepackage{graphicx}%
\usepackage{multirow}%
\usepackage{amsmath,amssymb,amsfonts}%
\usepackage{amsthm}%
\usepackage{mathrsfs}%
\usepackage[title]{appendix}%
\usepackage{xcolor}%
\usepackage{textcomp}%
\usepackage{manyfoot}%
\usepackage{booktabs}%
\usepackage{algorithm}%
\usepackage{algorithmicx}%
\usepackage{algpseudocode}%
\usepackage{listings}%
\usepackage{xltabular}
\usepackage{array}

\theoremstyle{thmstyleone}%
\theoremstyle{thmstyletwo}%
\newtheorem{remark}{Remark}%

\theoremstyle{thmstylethree}%

\begin{document}

\title[Bosonic and Fermionic Algebras from Deterministic Proper Time]{Universal Emergence of Bosonic and Fermionic Algebras in a Deterministic Proper-Time Framework}

\author*[1,2]{\fnm{Alessio} \sur{Maiezza}}\email{alessiomaiezza@gmail.com}

\affil*[1]{Dipartimento di Scienze Fisiche e Chimiche, Universit\`a degli Studi dell'Aquila, via Vetoio, I-67100, L'Aquila, Italy}

\affil[2]{INFN, Laboratori Nazionali del Gran Sasso, 67010 Assergi, L'Aquila, Italy}


\abstract{This work develops the deterministic, discrete proper-time framework introduced in our previous work, Eur.\ Phys.\ J.\ C \textbf{86} (2026) 829, in which quantum field theory emerges as an effective infrared description characterized by a running Planck constant. There, the effective quantization scale was inferred from the microscopic multiplicity unresolved by coarse-graining. Here, we provide its dynamical and operatorial realization and extend the construction to fermionic degrees of freedom. First, consistency under changes of macroscopic resolution leads to the structure of the Renormalization Group Equation, while the running Planck constant governs the crossover toward the deterministic regime. Second, we represent the reversible microscopic dynamics through finite-difference translations in field-configuration space. After coarse-graining, the resulting operator-valued canonical commutation relations reproduce the same quantization scale previously obtained from statistical microstate counting, thereby linking microscopic evolution to the emergent canonical algebra. Third, representing the same update within a Grassmann algebra yields the corresponding canonical anticommutation relations. The bosonic and fermionic sectors thus inherit a common effective Planck constant without introducing an independent fermionic update or quantization scale. Finally, we discuss possible implications for high-energy loop amplitudes and effective Hawking radiation.}




\maketitle

\section{Introduction}

One recurrent idea in approaches to a completion of the theory of gravitation is the introduction of a fundamental energy scale~\cite{Mead:1964zz,GARAY_1995,Kempf_1995,Padmanabhan:1996ap,Modesto_2009,Nicolini_2011,Bosso_2023,Bosso_2024,DAgostino:2025axy,Maiezza:2026wrp}. Related approaches rely on the generalized uncertainty principle~\cite{Maggiore:1993rv,Adler:1999bu,Ong:2018zqn,Ong:2023jkp,Bosso:2023aht,Ong:2025ent}, or on an effective or running Planck constant~\cite{adler2004quantum,Volovik:2009xs,Hossenfelder:2012uy}. A more radical possibility is to address the problem at the foundational level by assuming that nature is fundamentally deterministic~\cite{tHooft:2001qty,Palmer:2008jh,Hall:2010zzf,Hossenfelder:2019shy,Donadi:2020aqz,tHooft:2020qfg,tHooft:2020tuu,Powers:2021rfg,Hance:2022juc,Palmer:2023vfw,Arroyo:2024saq}. Ref.~\cite{Maiezza:2026wke} recently proposed a framework connecting these different perspectives. The construction builds on the proper-time formulation developed in Ref.~\cite{Maiezza:2026wrp}, where the introduction of a minimal proper time inversely proportional to the Planck mass, $\tau_{\min}\propto M_{\mathrm{Pl}}^{-1}$, leads to a controlled violation of unitarity in the deep ultraviolet, at energies of the order of the Planck mass $M_{\mathrm{Pl}}$, together with an effective running Planck constant, $\hbar_{\mathrm{eff}}$.\footnote{The loss of unitarity induces decoherence in the deep ultraviolet. Related decoherence mechanisms have also been considered in the literature; see, for example, Ref.~\cite{Petruzziello:2020wkd}.} The construction of Ref.~\cite{Maiezza:2026wke} starts from a deterministic and causally ordered set of events, from which quantum field theory emerges as an effective infrared description. As the fundamental scale is approached, $\hbar_{\mathrm{eff}}$ progressively vanishes and the quantum description, including its unitary evolution, reaches the boundary of its validity. In the infrared, the intrinsically discrete framework is also consistent with general relativity, with the spacetime manifold interpreted as an effective description.

The present work develops the deterministic coarse-graining framework introduced in our previous analysis~\cite{Maiezza:2026wke}. There, the microscopic evolution is described by an information-preserving discrete map, with each elementary update associated with the minimal proper-time interval. Continuous quantum field theory emerges only after coarse-graining over the underlying deterministic histories. The first aim of the present work is to clarify the relation between coarse-graining and the Renormalization Group Equation (RGE) of the emergent QFT. At energy scales $E\ll M_{\mathrm{Pl}}$, where $\hbar_{\mathrm{eff}}\approx\hbar$, changes in macroscopic resolution lead to the usual renormalization-group structure. As the Planck scale is approached, the continuum QFT description reaches the boundary of its validity and is progressively replaced by the underlying deterministic evolution. In this regime, the running of $\hbar_{\mathrm{eff}}$ describes the crossover toward $\hbar_{\mathrm{eff}}\to 0$. The standard renormalization-group flow and the running of the effective Planck constant therefore refer to two different regimes of the same change in resolution.

The second aim is to provide the missing dynamical and operatorial realization of the running of $\hbar_{\mathrm{eff}}$. In Ref.~\cite{Maiezza:2026wke}, the effective quantization scale was inferred from the unresolved microscopic multiplicity of the coarse-grained equivalence classes, providing a statistical interpretation of the gradual emergence of quantum behavior. The relation between the finite microscopic updates and the infinitesimal functional translations of the emergent field theory, however, was not made explicit. Since the fundamental dynamics proceeds through finite updates separated by $\tau_{\min}$, infinitesimal field translations should be regarded as an effective infrared description rather than as fundamental operations. We implement this idea in the bosonic sector by locally representing an elementary reversible update through finite translations in field-configuration space. The resulting finite-difference structure modifies both the Leibniz rule and the field-momentum commutator. The ordinary functional derivative and the canonical commutation relations are recovered when the coarse-grained states no longer resolve the individual microscopic update. After projection onto the macroscopic equivalence classes, the operator-valued deformation reproduces the same effective quantization scale previously obtained from statistical microstate counting. This agreement establishes the operatorial link between the reversible microscopic evolution and the emergent canonical algebra, rather than introducing a further prescription for the quantization scale.

The third aim, and the central new element of the present analysis, is the extension to fermionic degrees of freedom. In Ref.~\cite{Maiezza:2026wke}, the inclusion of microscopic variables associated with emergent fermionic fields was identified as an open direction toward a more realistic quantum field theory. Here, we take a first explicit step in this direction by constructing the corresponding canonical fermionic sector. The fermionic extension is based on the same deterministic update that underlies the bosonic construction. The fundamental map acts on the complete microscopic configuration and therefore also on the variables whose coarse-grained description gives rise to fermionic fields. Its effective action is represented within a Grassmann algebra, where the graded Leibniz rule leads to an operator-valued canonical anticommutator. The fermionic deformation is thus neither introduced independently nor postulated as a modification of the Berezin derivative, but inherits the same finite-update structure as the bosonic deformation. The two sectors share a common microscopic origin while retaining different algebraic realizations. For commuting field variables, the finite update gives rise to a deformed commutator; in the graded representation, it gives rise to a deformed anticommutator. After coarse-graining, the two canonical structures acquire the same normalization and are governed by the same effective Planck constant, $\hbar_{\mathrm{eff}}$. This common normalization is not imposed through an independent fermionic update or quantization scale, but follows from representing and projecting the same microscopic dynamics in the two canonical sectors. Within the common microscopic update and coarse-graining construction considered here, the effective quantization scale is therefore determined by the amount of microscopic information that remains unresolved, rather than by the bosonic or fermionic character of the emergent field. The fermionic construction consequently provides both a nontrivial consistency test of the framework and a necessary kinematical ingredient for its extension toward more realistic quantum field theories.

Finally, we provide an exploratory discussion of possible signatures of the deterministic crossover. In particular, a running effective Planck constant may affect the relative contribution of loop and tree-level amplitudes at high energies. A further possible implication is a modification of effective Hawking radiation. These observations are intended to identify phenomenological directions rather than to provide complete quantitative predictions.

The article is organized as follows. In Sec.~\ref{sec:previous-framework}, we briefly review the main ingredients of Ref.~\cite{Maiezza:2026wke} required for the present analysis. In Sec.~\ref{sec:emergent_rge_pregeom}, we discuss the relation between coarse-graining and the Renormalization Group Equation. In Sec.~\ref{sec:finite-difference-CCR}, we derive the bosonic canonical algebra from finite translations and recover the same effective Planck constant previously obtained from statistical microstate counting. In Sec.~\ref{sec:finite-CAR}, we extend the construction to fermionic degrees of freedom and show that the bosonic and fermionic canonical sectors inherit a common effective Planck constant from the same microscopic update and coarse-graining map. In Sec.~\ref{sec:phenomenological-signatures}, we discuss possible phenomenological signatures of the deterministic crossover. Sec.~\ref{end} provides the outlook.

\section{Deterministic emergence and effective quantization}
\label{sec:previous-framework}

In Ref.~\cite{Maiezza:2026wke}, we proposed a deterministic microscopic
counterpart of the proper-time formulation of quantum field theory.
The starting point was to consider, at the most fundamental level,
an ordered set of fundamental events:
\begin{equation}
\label{event_order}
\mathcal{E}^c
=
\{e_0^c,e_1^c,e_2^c,\ldots\},
\qquad
e_i^c \prec e_j^c
\quad\text{for}\quad
i<j \, ,
\end{equation}
where the superscript $c$ labels one of the possible causal chains\footnote{The use of discrete causally
ordered events has a structural analogy with causal set theory, where the fundamental spacetime 
structure is described by a locally finite partially ordered set~\cite{Bombelli:1987aa}. 
See also Ref.~\cite{Johnston:2010su} for a field-theoretic construction on causal sets. 
The present framework is nevertheless different: the chains support a deterministic proper-time update, while the continuous field description emerges through the specific coarse-graining introduced in Ref.~\cite{Maiezza:2026wke}.}.
The ontological microscopic data are the elementary events, their
causal relations, and the configurations $\phi(e_n^c)$ carried by
them. The events provide the causal support of the construction,
whereas $\phi(e_n^c)$ specifies the microscopic state on which the
deterministic evolution acts. Neither the events nor their
configurations are assumed to be spacetime points or quantum fields
defined on a pre-existing continuous manifold. The continuous field
$\phi(x)$ and its operator representation belong only to the
coarse-grained description.
The multiplicity of these chains is an essential ingredient of the
construction. A single chain would describe only an ordered sequence
of configurations and would not be sufficient to recover a field
theory. In the coarse-grained infrared description, a sufficiently
large set of chains provides the microscopic structure from which the
continuous spatial dependence of the effective field emerges,
schematically,
\begin{equation}
\phi(e_i^c)
\longrightarrow
\phi(x) \, .
\label{chain-to-field-summary}
\end{equation}
Thus, the chain label is resolved in the microscopic description,
while its information is absorbed into the continuous spacetime
dependence of the effective field.

The same limiting procedure underlies the emergence of the
relativistic spacetime description discussed in
Ref.~\cite{Maiezza:2026wke}. A single microscopic event is not
identified with a spacetime point, and no continuous metric is
assigned to an individual causal chain. Rather, the continuous
manifold and its metric description arise collectively from the
coarse-grained representation of a large network of interwoven
microscopic chains. In this representation, correlations among the
microscopic events acquire a smooth description in terms of
correlation functions of fields defined on the emergent spacetime.

Lorentz invariance is consequently an infrared property of the
emergent continuous theory, rather than a symmetry imposed on each
individual microscopic update. It is recovered in the ideal
continuum limit in which the discrete proper-time intervals and the
individual chain data are not resolved. At finite microscopic
resolution, corrections associated with the discrete event structure
are in general not excluded, so that Lorentz invariance should be
regarded as approximate away from this limit---see \cite{Maiezza:2026wke}. The form and magnitude
of such corrections are not universal within the present
model-independent construction and require a specific realization of
the microscopic map $F$.

At the microscopic level, proper-time evolution proceeds through
elementary updates separated by the nonzero interval
$\tau_{\min}$. Introducing the discrete proper-time label
\begin{equation}
\tau_s
=
s\,\tau_{\min} \, ,
\label{discrete-proper-time}
\end{equation}
the microscopic field configuration evolves according to the
information-preserving map
\begin{equation}
F\bigl(\phi(e_n^c)\bigr)
=
\phi(e_{n+1}^{c'})
\,,
\qquad
\phi_{s+1}
=
F(\phi_s) \, ,
\qquad
c'\neq c \, .
\label{deterministic-map-summary}
\end{equation}
The change of chain label is part of the microscopic update and does
not represent a probabilistic branching. For each microscopic
configuration, the deterministic map $F$ selects a unique successor.
Each application of $F$ therefore represents the finite change of the
complete microscopic configuration during one elementary proper-time
interval. At this level, neither continuous proper-time evolution nor
a canonical quantum algebra is assumed. Both belong to the effective
description obtained after coarse-graining over the discrete
microscopic dynamics and the underlying set of causal chains.

The microscopic evolution can be represented by the deterministic
transition kernel
\begin{equation}
K(\phi',\phi)
=
\delta_{\phi',F(\phi)} \, ,
\label{deterministic-kernel-summary}
\end{equation}
which assigns a unique successor to each microscopic configuration.
Since the microscopic dynamics is information-preserving, $F$ is
invertible on the microscopic configuration space. The elementary
forward and backward updates are consequently represented by $F$ and
$F^{-1}$, respectively. This reversible pair will provide the
microscopic origin of the symmetric finite translations introduced
below.

The macroscopic description is obtained by grouping microscopic
configurations into equivalence classes.
More precisely, the coarse-graining map acts on the microscopic
configurations generated by the deterministic evolution on the
underlying network of causal chains. Configurations that differ in
their microscopic event, chain, or history data, but cannot be
distinguished at the chosen macroscopic resolution, are represented
by the same effective state. The coarse-graining is therefore not an
average over a fundamental stochastic ensemble. It is a loss of
resolution among distinct configurations and deterministic histories
whose detailed microscopic information is inaccessible to the
effective observer. A conceptually related event-based construction
has been developed in Ref.~\cite{powers2026eventcentricapproachmodeling},
where causal networks of events are represented through binary sequences
and an epistemic restriction on the accessible microscopic information
leads to an emergent statistical description. The two frameworks differ
in their microscopic implementations, but share the view that quantum statistics
need not be fundamental and may instead reflect restricted access to an underlying event-based structure.
Below, the label $\mu$ denotes the macroscopic resolution at which the
configurations are compared. Different values of $\mu$ correspond to
different partitions of the microscopic configuration space into
equivalence classes. At this stage, the framework does not require a
specific functional dependence of the class cardinality on $\mu$;
only its limiting behavior is used. A coarse resolution leaves many
microscopic configurations unresolved within the same macroscopic
class, whereas a resolution approaching the fundamental scale
distinguishes progressively fewer internal alternatives.

At a given resolution $\mu$, an equivalence class is denoted by
\begin{equation}
[\phi_\mu]
=
\left\{
\phi_{\mu,1},
\ldots,
\phi_{\mu,n_\mu}
\right\} \, ,
\label{equivalence-class-summary}
\end{equation}
where its individual elements cannot be resolved by the
coarse-grained observer. The integer $n_\mu$ denotes the cardinality
of the class and therefore measures the number of microscopic
configurations represented by the same macroscopic state:
\begin{equation}
n_\mu
:=
|[\phi_\mu]|
\in
\mathbb{N} \, .
\end{equation}
Although the microscopic evolution is deterministic and
information-preserving, its projection onto the equivalence classes
does not, in general, preserve the full microscopic information and
therefore need not generate an exactly unitary effective evolution.
Part of the deterministic flow can cross the boundary between two
adjacent classes. In the minimal counting model considered in
Ref.~\cite{Maiezza:2026wke}, one microscopic configuration crosses the
boundary during each elementary update,
\begin{equation}
\Delta n_\mu
=
1 \, .
\label{minimal-crossing-summary}
\end{equation}
The fraction of microscopic configurations resolved as an outgoing
boundary transition is consequently
\begin{equation}
\eta_\mu
=
\frac{1}{n_\mu} \, ,
\label{resolved-fraction-summary}
\end{equation}
whereas the complementary fraction
\begin{equation}
Q(n_\mu)
=
1-\eta_\mu
=
1-\frac{1}{n_\mu}
\label{unresolved-fraction-summary}
\end{equation}
measures the fraction of microscopic configurations that remains
unresolved within the class.

The continuous quantum description arises only after coarse-graining
over the deterministic microscopic histories. In the extended
proper-time formulation, the corresponding macroscopic state is
described by a wave functional $\Psi[\phi,t,\tau]$, whose effective
evolution takes the Nambu--Schr\"odinger form~\cite{Maiezza:2026wrp}
\begin{equation}
i
\frac{\partial}{\partial\tau}
\Psi[\phi,t,\tau]
=
\hat H'
\Psi[\phi,t,\tau] \, .
\label{Nambu-Schrodinger-summary}
\end{equation}
Here, $\tau$ is the effective proper-time evolution parameter, while
the coordinate time $t$ is treated as an argument of the extended
field configuration. The physical post-constraint state is obtained
by integrating over the proper-time domain~\cite{Maiezza:2026wrp},
\begin{equation}
\Psi_R[\phi,t]
=
\int_{\tau_{\min}}^{\infty}
d\tau\,
\Psi[\phi,t,\tau] \, ,
\label{proper-time-projection-summary}
\end{equation}
where $\tau_{\min}$ is the fundamental proper-time interval separating
two consecutive microscopic updates.

In Ref.~\cite{Maiezza:2026wke}, the projected evolution was shown to
weight the effective quantum sector by the unresolved fraction
$Q(n_\mu)$. Within the minimal counting model, this gives
\begin{equation}
\hbar_{\rm eff}(\mu)
=
\hbar Q(n_\mu)
=
\hbar
\left(
1-\frac{1}{n_\mu}
\right) \, .
\label{heff-summary}
\end{equation}
Thus, Eq.~\eqref{heff-summary} is not introduced as a phenomenological
interpolation in the present analysis. It is the coarse-grained
quantization scale obtained in Ref.~\cite{Maiezza:2026wke} from the
relative weight of the unresolved microscopic sector.

For $n_\mu=1$, the equivalence class contains a single microscopic
configuration. No unresolved internal alternatives are then present,
and the evolution reduces to a deterministic permutation. Accordingly,
\begin{equation}
n_\mu=1
\qquad\Longrightarrow\qquad
Q(n_\mu)=0
\qquad\Longrightarrow\qquad
\hbar_{\rm eff}=0 \, .
\label{deterministic-limit-summary}
\end{equation}
In the opposite limit,
\begin{equation}
n_\mu\longrightarrow\infty
\qquad\Longrightarrow\qquad
Q(n_\mu)\longrightarrow1
\qquad\Longrightarrow\qquad
\hbar_{\rm eff}\longrightarrow\hbar \, ,
\label{quantum-limit-summary}
\end{equation}
and the standard normalization of the quantum theory is recovered.
Together with the smooth continuum limit discussed below, this leads
to the ordinary unitary QFT description. For comparison, a qualitatively opposite limiting behavior has recently
been considered in \cite{volovik2026quantumclassicalmechanicsvs}, where the pre-geometric
QFT regime is associated with $1/\hbar\to0$, or equivalently $\hbar\to\infty$,
rather than with the deterministic $\hbar_{\rm eff}\to0$ limit obtained here.

The effective canonical commutation relations were correspondingly
written as
\begin{equation}
\left[
\hat\phi(\vec{x}),
\hat\pi(\vec{y})
\right]_{\rm eff}
=
i\hbar_{\rm eff}(\mu)
\delta(\vec{x}-\vec{y}) \, .
\label{effective-CCR-summary}
\end{equation}
Equation~\eqref{effective-CCR-summary} expresses the normalization of
the canonical algebra by the fraction of microscopic information that
remains unresolved under coarse-graining. In particular, the
commutator vanishes in the deterministic limit and approaches the
standard canonical relation as the cardinality of the equivalence
classes grows.

The construction of Ref.~\cite{Maiezza:2026wke} therefore explains the
statistical and dynamical origin of the scalar factor multiplying the
effective canonical algebra. It does not, however, provide an explicit
representation of the canonical momentum directly in terms of the
elementary microscopic update.

This distinction is important. At the fundamental level, evolution
does not proceed through an infinitesimal variation of the field
configuration. During one proper-time interval $\tau_{\min}$, the
configuration is mapped by the finite transformation
\begin{equation}
\phi_s
\longmapsto
F(\phi_s) \, .
\label{finite-fundamental-update}
\end{equation}
The functional derivative entering the continuum momentum operator,
\begin{equation}
\hat\pi(\vec{x})
=
-i\hbar
\frac{\delta}{\delta\phi(\vec{x})} \, ,
\label{functional-momentum-summary}
\end{equation}
must therefore be understood as an effective smooth representation of
these finite microscopic changes rather than as a fundamental
operation \footnote{A distinct finite-proper-time construction has recently been developed in Ref.~\cite{Bakr:2026ngx}, where a retained endpoint is assigned to complete internal histories through a spectral operator calculus. Unlike the present framework, that endpoint is held fixed and is not introduced through coarse-graining over deterministic microscopic histories or through a deformation of the canonical algebra.}.

The purpose of the present work is to make this algebraic transition
explicit. We first represent an elementary reversible update locally
in field-configuration space by a pair of finite translations. The
forward and backward shifts provide a local coordinate realization of
the microscopic maps $F$ and $F^{-1}$, respectively. Their symmetric
combination defines a finite-difference operator, which reduces to the
ordinary functional derivative whenever the coarse-grained wave
functional is smooth on the scale of an elementary update.

This local translation representation does not require the global map
$F$ to act as a uniform translation over the entire configuration
space. Rather, it provides the minimal local realization of a finite
information-preserving update along a chosen configuration-space
coordinate. Within this realization, the finite parameter $\epsilon$
measures the displacement in field space associated with one
elementary update. It should not be identified with the spatial
lattice spacing, with a spacetime length, or with an additional
universal cutoff.

As we shall show, this representation leads simultaneously to a
deformed Leibniz rule and to an operator-valued deformation of the
field-momentum commutator. In the smooth coarse-grained regime, both
reduce to their ordinary continuum forms.

This algebraic construction is distinct from, although based on the
same microscopic framework as, the counting argument leading to
Eq.~\eqref{heff-summary}. The latter determines the effective strength
of quantization from the unresolved microscopic multiplicity. The
analysis below instead determines the algebra generated by a local
finite representation of the elementary microscopic update. Only
after this algebra has been obtained will its operator-valued
deformation be projected onto the coarse-grained equivalence classes.

\section{Renormalization Group from coarse-graining}
\label{sec:emergent_rge_pregeom}

The transition from the underlying deterministic substrate to an effective
quantum field theory (QFT) is tied to the emergence of the spacetime manifold
itself. At the fundamental level, the degrees of freedom are defined on a
discrete, pre-geometric set of causally ordered event chains. Let
$\phi(e_i^c)$ denote a microscopic variable evaluated at the event $e_i^c$ of
the chain $c$.

Continuous spacetime coordinates $x^\mu$ are not fundamental entities. They
emerge only when the observational resolution is much larger than the minimal
proper-time interval $\tau_{\min}$. In particular, if $t$ denotes the
characteristic proper-time duration involved in a macroscopic observation,
the continuum regime corresponds to
\begin{equation}
\frac{t}{\tau_{\min}}
\gg
1 \, .
\label{eq:qft_regime}
\end{equation}
In this regime, each macroscopic interval contains a large number of
elementary deterministic updates, and the individual steps of the microscopic
evolution are no longer resolved.

The microscopic dynamics unfolds in the extended space that includes the
proper-time coordinate $\tau$. The emergence of the standard four-dimensional
spacetime is governed by the Nambu-like constraint, which selects the physical
trajectories and projects the extended dynamics onto the four-dimensional
sector. Under this constraint, and after coarse-graining over a sufficiently
large number of elementary updates and causal chains, the discrete
pre-geometric structure admits an effective continuous representation,
schematically,
\begin{equation}
\left\{\mathcal{E}^c\right\}_{c}
\xrightarrow{\,t/\tau_{\min}\to\infty\,}
\mathcal{M} \, ,
\qquad
\phi(e_i^c)
\longrightarrow
\phi(x) \, ,
\label{eq:continuum_mapping}
\end{equation}
where $\mathcal{M}$ denotes the emergent spacetime manifold. This relation
does not imply a one-to-one identification between an individual microscopic
event and a spacetime point. Rather, the continuous field dependence arises
from the coarse-grained representation of the full set of microscopic chains.

Correspondingly, correlation functions defined on the microscopic event
network acquire a continuous infrared representation. For the two-point
function, this transition may be written schematically as
\begin{equation}
\langle 0|
\phi(e_i^c)\phi(e_j^{c'})
|0\rangle
\quad\longrightarrow\quad
\langle 0|
\phi(x)\phi(y)
|0\rangle \, .
\label{eq:twopoint_running}
\end{equation}
The states and correlation functions appearing on the left-hand side are
understood here in the linear representation of the deterministic microscopic
dynamics. Their use does not imply that the event variables are fundamental
quantum fields. In the following, the chain labels will be suppressed whenever
they are not needed explicitly.

The possibility of changing scale has a direct microscopic meaning in this
framework. Let $\ell_\mu$ denote the characteristic resolution length or
proper-time scale of the effective observer, with
\begin{equation}
\ell_\mu
\sim
\mu^{-1} \, ,
\label{eq:resolution_scale}
\end{equation}
in units in which $c=1$. The dimensionless ratio
\begin{equation}
N_\mu
=
\frac{\ell_\mu}{\tau_{\min}}
\label{eq:number_updates_resolution}
\end{equation}
measures the number of elementary proper-time intervals contained in the
effective resolution scale. This quantity should not be identified directly
with the cardinality $n_\mu$ of a coarse-grained equivalence class.
The former counts elementary intervals within the macroscopic resolution,
whereas the latter counts microscopic configurations represented by the same
macroscopic state. Their precise relation depends on the coarse-graining map.

The construction requires only that these quantities have compatible limiting
behavior. A resolution containing a large number of unresolved elementary
updates corresponds to equivalence classes with a large unresolved
microscopic multiplicity. Conversely, as the resolution approaches the
fundamental scale, the number of unresolved updates and the internal
multiplicity of the corresponding classes both decrease. Thus, no equality
between $N_\mu$ and $n_\mu$ is assumed, but both characterize complementary
aspects of the same loss of microscopic resolution.

A dilation of the macroscopic resolution,
\begin{equation}
\ell_\mu
\longrightarrow
\lambda\ell_\mu \, ,
\qquad
\mu
\longrightarrow
\lambda^{-1}\mu \, ,
\label{eq:dilation_resolution}
\end{equation}
therefore changes $N_\mu$ while leaving the fundamental interval
$\tau_{\min}$ fixed,
\begin{equation}
N_\mu
\longrightarrow
\lambda N_\mu \, ,
\qquad
\tau_{\min}
\longrightarrow
\tau_{\min} \, .
\label{eq:dilation_updates}
\end{equation}
Scale transformations in the emergent theory consequently have a microscopic
counterpart: they compare effective descriptions that leave different numbers
of elementary deterministic updates unresolved. In this sense, the change of
scale is intrinsic to the coarse-graining of the discrete dynamics rather
than being imposed only as a formal operation on an already continuous
theory.

Two distinct scale regimes must then be separated. For
\begin{equation}
N_\mu
\gg
1 \, ,
\qquad\text{equivalently}\qquad
\mu\tau_{\min}
\ll
1 \, ,
\label{eq:large_number_updates}
\end{equation}
the individual microscopic updates are unresolved and the continuous QFT
description is applicable. Changes of scale are then governed by the ordinary
renormalization-group flow of the effective couplings, fields, and correlation
functions. By the compatible limiting behavior of the coarse-graining map,
the corresponding equivalence classes contain a large unresolved
multiplicity, so that
\begin{equation}
\hbar_{\rm eff}
\simeq
\hbar \, .
\label{eq:hbar_ir_rge}
\end{equation}
Conversely, when the observational resolution approaches the fundamental
proper-time interval,
\begin{equation}
N_\mu
=
\mathcal{O}(1) \, ,
\qquad\text{equivalently}\qquad
\mu\tau_{\min}
=
\mathcal{O}(1) \, ,
\label{eq:fundamental_resolution}
\end{equation}
the continuum QFT description approaches the boundary of its validity. The
ordinary renormalization-group flow then no longer provides a complete
description of the change of scale. In this regime, the varying microscopic
multiplicity also changes the normalization of the canonical algebra through
the effective Planck constant $\hbar_{\rm eff}$.

The two types of scale dependence are therefore distinct, although they arise
from the same underlying change of microscopic resolution. The standard
renormalization-group flow describes the scale dependence of couplings,
fields, and correlation functions within the emergent continuous QFT. The
running of $\hbar_{\rm eff}$ instead describes the crossover in which the
continuous quantum representation progressively gives way to the underlying
deterministic dynamics. The latter should not be interpreted as an additional
beta function within an otherwise fundamental continuum QFT.

To make the first regime explicit, consider the two-point function of the
emergent field. On the microscopic event network, let
\begin{equation}
G_0^{(2)}
\bigl(e_i^c,e_j^{c'};\tau_{\min},g_0\bigr)
=
\langle 0|
\phi_0(e_i^c)\phi_0(e_j^{c'})
|0\rangle \, ,
\label{eq:bare_green}
\end{equation}
where $g_0$ denotes the microscopic, or bare, parameters in the linear
representation of the deterministic dynamics. The subscript $0$
distinguishes these quantities from their renormalized continuum
counterparts.

After coarse-graining into the regime $N_\mu\gg1$, the microscopic correlator
admits a continuous representation at the resolution scale $\mu$. The
corresponding effective field is related to the coarse-grained microscopic
field representation through
\begin{equation}
\phi(x;\mu)
=
Z^{-1/2}(\mu,\tau_{\min},g_0)\,
\phi_{0,\mu}(x) \, ,
\label{eq:field_renormalization}
\end{equation}
where $\phi_{0,\mu}(x)$ denotes the continuous representation, at resolution
$\mu$, of the microscopic event variables. The dependence of
$\phi_{0,\mu}(x)$ on $\mu$ belongs to the coarse-graining map and should be
distinguished from the microscopic dynamics itself, which remains fixed.

The continuous two-point function is consequently related to the
coarse-grained representation of the microscopic correlator by
\begin{equation}
G^{(2)}(x,y;\mu,g)
=
Z^{-1}(\mu,\tau_{\min},g_0)\,
G_{0,\mu}^{(2)}(x,y;\tau_{\min},g_0) \, ,
\label{eq:green_relation}
\end{equation}
where $g=g(\mu)$ is the renormalized coupling characterizing the effective
continuous description. The quantity $G_{0,\mu}^{(2)}$ is the representation
at scale $\mu$ of the same underlying microscopic correlator
$G_0^{(2)}(e_i^c,e_j^{c'};\tau_{\min},g_0)$.

The scale $\mu$ labels the resolution at which the emergent continuous theory
is parametrized. It does not modify the underlying deterministic map, the
microscopic parameters, or the fundamental proper-time interval. Therefore,
at fixed $\tau_{\min}$ and fixed bare parameters, the microscopic correlator
defined on the event network is independent of $\mu$,
\begin{equation}
\left.
\mu\frac{d}{d\mu}
G_0^{(2)}
\bigl(e_i^c,e_j^{c'};\tau_{\min},g_0\bigr)
\right|_{\tau_{\min},g_0}
=
0 \, .
\label{eq:independence_principle}
\end{equation}
Once expressed in continuous variables, this condition states that different
choices of $\mu$ provide equivalent parametrizations of the same microscopic
correlator. The derivative is understood at fixed effective coordinates
$x$ and $y$. This is the emergent counterpart of the standard independence
of a bare correlation function from the arbitrary scale used to parametrize
its renormalized representation. In the present framework, the change of that
scale also compares coarse-grained descriptions containing different numbers
of unresolved elementary updates.

Using Eq.~\eqref{eq:green_relation}, the independence condition becomes
\begin{equation}
\left.
\mu\frac{d}{d\mu}
\left[
Z(\mu,\tau_{\min},g_0)\,
G^{(2)}(x,y;\mu,g)
\right]
\right|_{\tau_{\min},g_0}
=
0 \, .
\label{eq:cs_derivation}
\end{equation}
Here and below, Eq.~\eqref{eq:cs_derivation} is understood in the continuum
regime $\mu\tau_{\min}\ll1$. In this regime, corrections that explicitly
resolve the fundamental proper-time scale can be neglected, and the remaining
scale dependence is parametrized by the running coupling and the field
renormalization. For simplicity, we consider a massless model, or a regime in
which mass terms can be neglected, and a single effective coupling.

Expanding the total derivative then gives the Callan--Symanzik equation for
the emergent two-point function,
\begin{equation}
\left[
\mu\frac{\partial}{\partial\mu}
+
\beta(g)\frac{\partial}{\partial g}
+
2\gamma(g)
\right]
G^{(2)}(x,y;\mu,g)
=
0 \, ,
\label{eq:callan_symanzik}
\end{equation}
where
\begin{equation}
\beta(g)
=
\left.
\mu\frac{\partial g}{\partial\mu}
\right|_{\tau_{\min},g_0}
\label{eq:beta_function}
\end{equation}
and
\begin{equation}
\gamma(g)
=
\left.
\frac{1}{2}\mu
\frac{\partial\ln Z}{\partial\mu}
\right|_{\tau_{\min},g_0}
=
\left.
\frac{1}{2}\frac{\mu}{Z}
\frac{\partial Z}{\partial\mu}
\right|_{\tau_{\min},g_0}
\, .
\label{eq:anomalous_dimension}
\end{equation}
For an emergent $n$-point function, the corresponding
field-renormalization term is $n\gamma(g)$. The extension to several
couplings is obtained by replacing
$\beta(g)\partial/\partial g$ with the corresponding sum over running
couplings.

Equation~\eqref{eq:callan_symanzik} does not determine the explicit form of
$\beta(g)$ or $\gamma(g)$. These functions depend on the interactions and on
the continuous QFT obtained after coarse-graining. The formal derivation of
the Callan--Symanzik equation follows the standard independence of the
microscopic description from the sliding scale. The role of the
deterministic substrate is instead to provide a microscopic interpretation
of that scale variation and to identify the regime in which the continuous
renormalization-group description applies.

A scale transformation changes the number of elementary proper-time updates
that remain unresolved, while the microscopic interval $\tau_{\min}$, the
bare parameters, and the deterministic update remain fixed. When that number
is large, the comparison between resolutions is represented by the standard
Callan--Symanzik flow of the emergent QFT. When the resolution approaches the
scale of a single update, corrections depending explicitly on
$\mu\tau_{\min}$ can no longer be neglected. The assumptions underlying the
continuous correlation functions then cease to apply, and the scale
dependence is no longer exhausted by the running of $g$ and $Z$. It also
involves the deformation of the effective canonical algebra through
$\hbar_{\rm eff}$.

The complete scale structure may therefore be summarized as
\begin{equation}
\begin{aligned}
\ell_\mu \gg \tau_{\min}
&:
&
\text{continuous QFT}
&\quad\longrightarrow\quad
\beta(g),\ \gamma(g),\ Z(\mu) \, ,
\\[3pt]
\ell_\mu \sim \tau_{\min}
&:
&
\text{deterministic crossover}
&\quad\longrightarrow\quad
Q(n_\mu),\ \hbar_{\rm eff}(\mu) \, .
\end{aligned}
\label{eq:two_scale_flows}
\end{equation}
The first line describes the renormalization-group flow within the emergent
continuum theory. The second describes the change in the strength and
structure of the quantum representation itself. The standard
renormalization-group equation and the running of $\hbar_{\rm eff}$ are
therefore not competing descriptions of the same regime. They are successive
effective manifestations of the same underlying change of microscopic
resolution.

\section{Canonical algebra from finite translations in field space}
\label{sec:finite-difference-CCR}

Having established how macroscopic scale transformations reflect the varying
resolution of discrete microscopic updates, we now turn to the explicit construction
of the emergent canonical algebra. To understand how finite proper-time
steps modify the continuous field derivatives at a given scale, we first analyze
how an elementary reversible update is represented locally in configuration space.

Before addressing the functional field-theory construction, it is
useful to illustrate the underlying mechanism in ordinary
one-dimensional quantum mechanics. The following example should not
be interpreted as the microscopic dynamics itself, which is specified
by the map $F$. Its purpose is instead to display the local algebra
obtained when a reversible pair of elementary updates, $F$ and
$F^{-1}$, is represented in a configuration-space coordinate by the
finite shifts $T_\epsilon$ and $T_{-\epsilon}$.

Consider the symmetric finite-difference operator
\begin{equation}
D_\epsilon\psi(x)
=
\frac{
\psi(x+\epsilon)
-
\psi(x-\epsilon)
}{
2\epsilon
}
\label{QM-finite-difference}
\end{equation}
and define the corresponding momentum operator by
\begin{equation}
\hat p_\epsilon
=
-i\hbar D_\epsilon \, .
\label{QM-finite-momentum}
\end{equation}
A direct calculation gives
\begin{align}
\left[
\hat x,
\hat p_\epsilon
\right]\psi(x)
&=
\frac{i\hbar}{2}
\left[
\psi(x+\epsilon)
+
\psi(x-\epsilon)
\right]
\nonumber
\\[1mm]
&=
i\hbar C_\epsilon\psi(x) \, ,
\label{QM-finite-commutator-action}
\end{align}
where
\begin{equation}
C_\epsilon
=
\frac{
T_\epsilon+T_{-\epsilon}
}{2},
\qquad
T_{\pm\epsilon}\psi(x)
=
\psi(x\pm\epsilon) \, .
\label{QM-average-operator}
\end{equation}
Thus, the finite-difference momentum satisfies the exact deformed
commutation relation
\begin{equation}
\left[
\hat x,
\hat p_\epsilon
\right]
=
i\hbar C_\epsilon \, .
\label{QM-finite-CCR}
\end{equation}
The same averaging operator appears in the product rule. For arbitrary
functions $f(x)$ and $g(x)$,
\begin{equation}
D_\epsilon(fg)
=
\left(
C_\epsilon f
\right)
\left(
D_\epsilon g
\right)
+
\left(
D_\epsilon f
\right)
\left(
C_\epsilon g
\right) \, .
\label{QM-deformed-Leibniz}
\end{equation}
The deformation of the commutator is therefore a direct consequence
of the fact that a finite difference is not an ordinary derivation.
Indeed, since
\begin{equation}
D_\epsilon(x\psi)
=
xD_\epsilon\psi
+
C_\epsilon\psi \, ,
\label{QM-coordinate-product}
\end{equation}
multiplication by $-i\hbar$ immediately yields
Eq.~\eqref{QM-finite-CCR}.

In the smooth regime,
\begin{equation}
D_\epsilon
=
\frac{d}{dx}
+
\frac{\epsilon^2}{6}
\frac{d^3}{dx^3}
+
\mathcal{O}(\epsilon^4),
\qquad
C_\epsilon
=
1
+
\frac{\epsilon^2}{2}
\frac{d^2}{dx^2}
+
\mathcal{O}(\epsilon^4) \, ,
\label{QM-smooth-expansion}
\end{equation}
and therefore
\begin{equation}
\epsilon\frac{d}{dx}
\longrightarrow
0
\qquad\Longrightarrow\qquad
\left[
\hat x,
\hat p_\epsilon
\right]
\longrightarrow
i\hbar \, .
\label{QM-canonical-limit}
\end{equation}
Here the limit denotes a regime in which the relevant states do not
resolve the finite displacement $\epsilon$; it does not require the
fundamental update scale itself to vanish.

The same local algebra can be used to represent an elementary update
along one coordinate of the regularized field-configuration space.
The corresponding identifications are
\begin{equation}
x
\longrightarrow
\phi_j,
\qquad
D_\epsilon
\longrightarrow
D_{\epsilon,j},
\qquad
\hat p_\epsilon
\longrightarrow
\hat\pi_{\epsilon,j},
\qquad
C_\epsilon
\longrightarrow
C_{\epsilon,j} \, .
\label{QM-field-correspondence}
\end{equation}
Accordingly, the one-dimensional commutator is promoted to
\begin{equation}
\left[
\hat x,
\hat p_\epsilon
\right]
=
i\hbar C_\epsilon
\qquad
\longrightarrow
\qquad
\left[
\hat\phi_j,
\hat\pi_{\epsilon,k}
\right]
=
i\hbar
\delta_{jk}
C_{\epsilon,k} \, .
\label{QM-field-commutator-correspondence}
\end{equation}
The Kronecker delta appears because the finite-difference operator
$D_{\epsilon,k}$ acts along the $k$-th independent direction of
field-configuration space,
\begin{equation}
D_{\epsilon,k}\phi_j
=
\delta_{jk} \, .
\label{field-coordinate-finite-derivative}
\end{equation}
These identifications concern the local representation of the update,
not the global form of the deterministic map. A general map $F$ may
couple several field coordinates and generate
configuration-dependent displacements. The constant-step,
single-coordinate translation considered here is the minimal local
model needed to isolate the canonical algebra associated with one
field-space direction.

The field-theory derivation follows the same algebraic steps as the
one-dimensional example; the additional indices keep track of the
independent coordinates of the regularized configuration space. After
removal of the spatial regulator, $\delta_{jk}$ is replaced, with the
appropriate lattice normalization, by the spatial delta function
$\delta(\vec{x}-\vec{y})$.

We now implement the construction explicitly in field-configuration
space, without assuming the functional canonical commutation
relations. For definiteness, the field theory is first formulated on a
spatial lattice. A field configuration is represented by
\begin{equation}
\boldsymbol{\phi}
=
\left(
\phi_1,
\ldots,
\phi_N
\right) \, ,
\label{lattice-field-configuration}
\end{equation}
and the corresponding state by a wave functional
$\Psi(\boldsymbol{\phi})$. The spatial lattice is used only as a
regulator. In particular, the finite step introduced below is a
displacement in field-configuration space and should not be confused
with the spatial lattice spacing.

The deterministic map advances the complete microscopic configuration
by one proper-time step. To extract the local algebra associated with
the $j$-th configuration-space coordinate, we represent the
corresponding elementary reversible update by the pair of finite
translations
\begin{equation}
\hat T_{\pm\epsilon,j}
\Psi(\boldsymbol{\phi})
=
\Psi\left(
\boldsymbol{\phi}
\pm
\epsilon\boldsymbol{e}_j
\right) \, ,
\label{finite-field-translations}
\end{equation}
where $\boldsymbol{e}_j$ denotes the unit vector along the $j$-th
field coordinate. The forward and backward translations provide the
local coordinate representation of $F$ and $F^{-1}$ along the chosen
field-space direction. The parameter $\epsilon$ denotes the
field-space displacement resolved in one elementary update within
this local model. It is neither the spatial lattice spacing nor, in
general, a spacetime length.

The use of a constant $\epsilon$ and of a single direction
$\boldsymbol e_j$ defines the minimal local realization. A general
microscopic update may involve several field coordinates and
configuration-dependent displacements, but these additional features
are not required for the algebraic mechanism studied here.

The symmetric finite-difference operator is
\begin{equation}
\hat D_{\epsilon,j}
=
\frac{
\hat T_{\epsilon,j}
-
\hat T_{-\epsilon,j}
}{
2\epsilon
} \, ,
\label{finite-field-derivative}
\end{equation}
while the associated symmetric averaging operator is
\begin{equation}
\hat C_{\epsilon,j}
=
\frac{
\hat T_{\epsilon,j}
+
\hat T_{-\epsilon,j}
}{
2
} \, .
\label{finite-field-average}
\end{equation}
A finite-difference momentum operator can then be defined by
\begin{equation}
\hat\pi_{\epsilon,j}
=
-i\hbar
\hat D_{\epsilon,j} \, .
\label{finite-field-momentum}
\end{equation}
On a domain invariant under the translations and with the usual
conditions ensuring
$\hat T_{\epsilon,j}^{\dagger}=\hat T_{-\epsilon,j}$,
the operator $\hat D_{\epsilon,j}$ is anti-Hermitian, while
$\hat\pi_{\epsilon,j}$ and $\hat C_{\epsilon,j}$ are Hermitian.

Acting on an arbitrary wave functional, one finds
\begin{align}
\hat\phi_j
\hat\pi_{\epsilon,k}
\Psi(\boldsymbol{\phi})
&=
-\frac{i\hbar\phi_j}{2\epsilon}
\left[
\Psi\left(
\boldsymbol{\phi}
+
\epsilon\boldsymbol{e}_k
\right)
-
\Psi\left(
\boldsymbol{\phi}
-
\epsilon\boldsymbol{e}_k
\right)
\right] \, ,
\label{phi-pi-action}
\\[1mm]
\hat\pi_{\epsilon,k}
\hat\phi_j
\Psi(\boldsymbol{\phi})
&=
-\frac{i\hbar}{2\epsilon}
\left[
\left(
\phi_j+\epsilon\delta_{jk}
\right)
\Psi\left(
\boldsymbol{\phi}
+
\epsilon\boldsymbol{e}_k
\right)
\right.
\nonumber
\\
&\hspace{31mm}
\left.
-
\left(
\phi_j-\epsilon\delta_{jk}
\right)
\Psi\left(
\boldsymbol{\phi}
-
\epsilon\boldsymbol{e}_k
\right)
\right] \, .
\label{pi-phi-action}
\end{align}
Taking the difference gives
\begin{align}
\left[
\hat\phi_j,
\hat\pi_{\epsilon,k}
\right]
\Psi(\boldsymbol{\phi})
&=
\frac{i\hbar}{2}
\delta_{jk}
\left[
\Psi\left(
\boldsymbol{\phi}
+
\epsilon\boldsymbol{e}_k
\right)
+
\Psi\left(
\boldsymbol{\phi}
-
\epsilon\boldsymbol{e}_k
\right)
\right]
\nonumber
\\[1mm]
&=
i\hbar
\delta_{jk}
\hat C_{\epsilon,k}
\Psi(\boldsymbol{\phi}) \, .
\label{finite-CCR-action}
\end{align}
The finite-translation algebra therefore satisfies the exact
commutation relation
\begin{equation}
\left[
\hat\phi_j,
\hat\pi_{\epsilon,k}
\right]
=
i\hbar
\delta_{jk}
\hat C_{\epsilon,k}
\, .
\label{finite-CCR}
\end{equation}
Unlike the canonical commutation relation, the right-hand side of
Eq.~\eqref{finite-CCR} is not proportional to the identity operator.
The deformation is controlled by the symmetric finite translation
$\hat C_{\epsilon,k}$ and is therefore operator-valued.

This deformation is directly related to the fact that a finite
difference is not an ordinary derivation. Let
$A(\boldsymbol{\phi})$ and $B(\boldsymbol{\phi})$ be two arbitrary
functionals. From the definitions
\eqref{finite-field-derivative} and
\eqref{finite-field-average}, one obtains
\begin{equation}
\hat D_{\epsilon,j}(AB)
=
\left(
\hat C_{\epsilon,j}A
\right)
\left(
\hat D_{\epsilon,j}B
\right)
+
\left(
\hat D_{\epsilon,j}A
\right)
\left(
\hat C_{\epsilon,j}B
\right)
\, .
\label{deformed-Leibniz}
\end{equation}
Equation~\eqref{deformed-Leibniz} is the exact Leibniz rule associated
with the symmetric finite difference. The same averaging operator
$\hat C_{\epsilon,j}$ controls both the product rule and the
field-momentum commutator. Consequently, the deformation of the
canonical algebra is not introduced independently, but follows from
the finite representation of the elementary reversible update in
field-configuration space.

The relation between the two deformations can be displayed directly.
Since
\begin{equation}
\hat D_{\epsilon,k}\phi_j
=
\delta_{jk}
\end{equation}
and
\begin{equation}
\hat C_{\epsilon,k}\phi_j
=
\phi_j \, ,
\end{equation}
the deformed Leibniz rule gives
\begin{align}
\hat D_{\epsilon,k}
\left(
\phi_j\Psi
\right)
&=
\phi_j
\hat D_{\epsilon,k}\Psi
+
\delta_{jk}
\hat C_{\epsilon,k}\Psi \, .
\label{Leibniz-coordinate}
\end{align}
Multiplication by $-i\hbar$ then reproduces
Eq.~\eqref{finite-CCR}. The deformed commutator and the deformed
Leibniz rule are therefore two manifestations of the same underlying
finite-update structure.

For wave functionals admitting a smooth representation, the
translation operators can be written formally as
\begin{equation}
\hat T_{\pm\epsilon,j}
=
\exp\left(
\pm\epsilon
\frac{\partial}{\partial\phi_j}
\right) \, ,
\label{translation-exponential}
\end{equation}
so that
\begin{align}
\hat D_{\epsilon,j}
&=
\frac{1}{\epsilon}
\sinh\left(
\epsilon
\frac{\partial}{\partial\phi_j}
\right)
\nonumber
\\
&=
\frac{\partial}{\partial\phi_j}
+
\frac{\epsilon^2}{6}
\frac{\partial^3}{\partial\phi_j^3}
+
\mathcal{O}(\epsilon^4) \, ,
\label{D-expansion}
\\[2mm]
\hat C_{\epsilon,j}
&=
\cosh\left(
\epsilon
\frac{\partial}{\partial\phi_j}
\right)
\nonumber
\\
&=
1
+
\frac{\epsilon^2}{2}
\frac{\partial^2}{\partial\phi_j^2}
+
\mathcal{O}(\epsilon^4) \, .
\label{C-expansion}
\end{align}
It follows that
\begin{equation}
\left[
\hat\phi_j,
\hat\pi_{\epsilon,k}
\right]
=
i\hbar
\delta_{jk}
\left[
1
+
\frac{\epsilon^2}{2}
\frac{\partial^2}{\partial\phi_k^2}
+
\mathcal{O}(\epsilon^4)
\right] \, .
\label{finite-CCR-expansion}
\end{equation}
The ordinary lattice canonical algebra is recovered in the
coarse-grained sector whose wave functionals do not resolve the
field-space displacement associated with one elementary update. More
precisely, the relevant smoothness condition is
\begin{equation}
\left|
\epsilon
\frac{\partial\Psi}{\partial\phi_j}
\right|
\ll
|\Psi| \, ,
\label{wavefunctional-smoothness}
\end{equation}
together with the analogous suppression of the higher derivatives
appearing in Eqs.~\eqref{D-expansion} and \eqref{C-expansion}.
Symbolically, this regime may be written as
\begin{equation}
\epsilon
\frac{\partial}{\partial\phi_j}
\longrightarrow
0 \, .
\label{smooth-field-space-limit}
\end{equation}
In this regime,
\begin{equation}
\hat D_{\epsilon,j}
\longrightarrow
\frac{\partial}{\partial\phi_j}
\, ,
\qquad
\hat C_{\epsilon,j}
\longrightarrow
1 \, ,
\label{finite-continuum-limit}
\end{equation}
and hence
\begin{equation}
\left[
\hat\phi_j,
\hat\pi_k
\right]
=
i\hbar\delta_{jk} \, .
\label{canonical-lattice-CCR}
\end{equation}
At the same time, Eq.~\eqref{deformed-Leibniz} reduces to the ordinary
Leibniz rule. The canonical commutation relations and the standard
product rule thus emerge in the same smooth-field-space regime.

This is an infrared smoothness limit, not a removal of the fundamental
proper-time interval. The microscopic evolution remains discrete,
with consecutive configurations separated by one update, while its
action becomes indistinguishable from an infinitesimal
configuration-space translation on sufficiently smooth macroscopic
states.

After the appropriate rescaling of the lattice variables and removal
of the spatial regulator, the continuum counterpart of
Eq.~\eqref{finite-CCR} may be written schematically as
\begin{equation}
\left[
\hat\phi(\vec{x}),
\hat\pi_\epsilon(\vec{y})
\right]
=
i\hbar
\delta(\vec{x}-\vec{y})
\hat C_{\epsilon,\vec{y}} \, ,
\label{functional-finite-CCR}
\end{equation}
where $\hat C_{\epsilon,\vec{y}}$ represents a symmetric finite
translation along the field coordinate associated with the spatial
point $\vec{y}$. Equation~\eqref{functional-finite-CCR} is to be
understood as the effective continuum representation of the spatially
regularized construction. It should not be interpreted as an
unregulated pointwise shift of a fundamental continuum field
configuration.

The operator identity obtained so far does not use the coarse-grained
counting summarized in Section~\ref{sec:previous-framework}.
Equation~\eqref{heff-summary} determines the scalar strength of
effective quantization from the unresolved multiplicity of the
equivalence classes. By contrast, Eq.~\eqref{finite-CCR} describes the
pre-canonical algebra associated with a local finite representation
of the elementary reversible update in field-configuration space.

\subsection{Matching to the coarse-grained quantization scale}
\label{subsec:finite-matching}

We now connect the operator-valued deformation generated by finite
translations in field-configuration space with the effective
quantization factor determined by the cardinality of the
coarse-grained equivalence classes.

Following the uniform linear representation adopted in
Ref.~\cite{Maiezza:2026wke}, let
\begin{equation}
[\phi_\mu]
=
\left\{
\phi\in\Lambda
\,\middle|\,
C_\mu(\phi)=\phi_\mu
\right\},
\qquad
n_\mu
=
\left|[\phi_\mu]\right| \, ,
\label{equivalence-class-matching}
\end{equation}
where $\Lambda$ is the set of all microstates $\phi(e_n)$, and let
\begin{equation}
\lvert\phi_\mu\rangle
=
\frac{1}{\sqrt{n_\mu}}
\sum_{\phi\in[\phi_\mu]}
\lvert\phi\rangle
\label{coarse-state-matching}
\end{equation}
be the normalized linear representative of the corresponding
macroscopic equivalence class.

Consider the finite translations introduced in
Section~\ref{sec:finite-difference-CCR},
\begin{equation}
\hat T_{\pm\epsilon}
\lvert\phi\rangle
=
\lvert T_{\pm\epsilon}\phi\rangle,
\qquad
\langle\phi'|
\hat T_{\pm\epsilon}
|\phi\rangle
=
\delta_{\phi',T_{\pm\epsilon}\phi} \, .
\label{microscopic-finite-shifts}
\end{equation}
Within the local realization used above, $T_\epsilon$ and
$T_{-\epsilon}=T_\epsilon^{-1}$ represent the forward and backward
elementary updates. For each direction, we denote by $b_\mu^\pm$ the
number of configurations that are translated outside the equivalence
class,
\begin{equation}
b_\mu^\pm
=
\left|
\left\{
\phi\in[\phi_\mu]
\,\middle|\,
T_{\pm\epsilon}\phi\notin[\phi_\mu]
\right\}
\right| \, .
\label{boundary-cardinalities}
\end{equation}
It follows that $n_\mu-b_\mu^\pm$ configurations are translated
internally.

Using Eqs.~\eqref{coarse-state-matching} and
\eqref{microscopic-finite-shifts}, the projection of each translation
onto the microscopic class gives
\begin{align}
\langle\phi_\mu|
\hat T_{\pm\epsilon}
|\phi_\mu\rangle
&=
\frac{1}{n_\mu}
\sum_{\phi,\phi'\in[\phi_\mu]}
\delta_{\phi',T_{\pm\epsilon}\phi}
\nonumber
\\
&=
1-\frac{b_\mu^\pm}{n_\mu} \, .
\label{projected-shift-expectation}
\end{align}
Thus, the projected matrix element measures the fraction of finite
updates that remain internal to the equivalence class.

Since $\hat T_{-\epsilon}=\hat T_\epsilon^{-1}$, every internal
forward transition between two configurations of the equivalence
class is paired with a unique internal backward transition. The
numbers of internal transitions are therefore equal in the two
directions. Since the class contains $n_\mu$ configurations, the
corresponding numbers of outgoing transitions must also coincide,
\begin{equation}
b_\mu^+
=
b_\mu^-
\equiv
b_\mu \, .
\label{reversible-boundary-equality}
\end{equation}
Reversibility implies equality of the two crossing numbers, not
necessarily identity of the configurations that cross the boundary
in the two directions.

For the symmetric translation operator
\begin{equation}
\hat C_\epsilon
=
\frac{
\hat T_{\epsilon}
+
\hat T_{-\epsilon}
}{2} \, ,
\label{symmetric-translation-matching}
\end{equation}
one then has
\begin{equation}
\langle\phi_\mu|
\hat C_\epsilon
|\phi_\mu\rangle
=
1-\frac{b_\mu}{n_\mu} \, .
\label{C-expectation-reversible}
\end{equation}

We now specialize to the minimal boundary-crossing realization,
\begin{equation}
b_\mu
=
1 \, .
\label{minimal-symmetric-crossing}
\end{equation}
Thus, the equality of the forward and backward crossing numbers
follows from the reversibility of the microscopic update, while
$b_\mu=1$ is the minimal counting assumption inherited from
Ref.~\cite{Maiezza:2026wke}. For either direction, one microscopic
configuration is resolved as a boundary transition, while the
remaining $n_\mu-1$ configurations are translated internally.
Equation~\eqref{C-expectation-reversible} then reduces to
\begin{equation}
\langle\phi_\mu|
\hat C_\epsilon
|\phi_\mu\rangle
=
1-\frac{1}{n_\mu}
=
Q(n_\mu)
\, .
\label{C-expectation-minimal}
\end{equation}
This reduction differs from the smooth-field-space limit discussed
above. The latter states that a finite shift acts approximately as the
identity on sufficiently smooth wave functionals, whereas
Eq.~\eqref{C-expectation-minimal} measures the fraction of microscopic
translations that remain internal to a coarse-grained equivalence
class. In the deep infrared, the two descriptions are compatible:
$n_\mu\gg1$ implies $Q(n_\mu)\simeq1$, while the elementary
field-space displacement is not resolved.

Defining the projector
\begin{equation}
\hat{\mathcal P}_\mu
=
\lvert\phi_\mu\rangle
\langle\phi_\mu\rvert \, ,
\label{macroscopic-state-projector}
\end{equation}
one has
\begin{equation}
\hat{\mathcal P}_\mu
\hat C_\epsilon
\hat{\mathcal P}_\mu
=
\langle\phi_\mu|
\hat C_\epsilon
|\phi_\mu\rangle
\hat{\mathcal P}_\mu \, .
\label{rank-one-C-projection}
\end{equation}
Therefore, using Eq.~\eqref{C-expectation-minimal}, we obtain
\begin{equation}
\hat{\mathcal P}_\mu
\hat C_\epsilon
\hat{\mathcal P}_\mu
=
Q(n_\mu)
\hat{\mathcal P}_\mu
=
\left(
1-\frac{1}{n_\mu}
\right)
\hat{\mathcal P}_\mu
\, .
\label{macroscopic-C-projection}
\end{equation}
The operator-valued deformation thus becomes a scalar upon restriction
to the macroscopic state representing the equivalence class.

Applying Eq.~\eqref{macroscopic-C-projection} to the finite-difference
commutator
\begin{equation}
\left[
\hat\phi_j,
\hat\pi_{\epsilon,k}
\right]
=
i\hbar
\delta_{jk}
\hat C_{\epsilon,k}
\label{finite-CCR-recalled}
\end{equation}
gives
\begin{align}
\hat{\mathcal P}_\mu
\left[
\hat\phi_j,
\hat\pi_{\epsilon,k}
\right]
\hat{\mathcal P}_\mu
&=
i\hbar
Q(n_\mu)
\delta_{jk}
\hat{\mathcal P}_\mu \, .
\label{projected-finite-CCR}
\end{align}
Equivalently, the corresponding matrix element in the normalized
macroscopic state is
\begin{equation}
\langle\phi_\mu|
\left[
\hat\phi_j,
\hat\pi_{\epsilon,k}
\right]
|\phi_\mu\rangle
=
i\hbar
Q(n_\mu)
\delta_{jk} \, .
\label{macroscopic-CCR-matrix-element}
\end{equation}
The effective canonical algebra is defined as the scalar relation that
reproduces these matrix elements in the coarse-grained macroscopic
sector. It is therefore
\begin{equation}
\left[
\hat\phi_j,
\hat\pi_k
\right]_{\rm eff}
=
i\hbar
\left(
1-\frac{1}{n_\mu}
\right)
\delta_{jk}
=
i\hbar_{\rm eff}(\mu)
\delta_{jk}
\, ,
\label{matched-effective-CCR}
\end{equation}
where
\begin{equation}
\hbar_{\rm eff}(\mu)
=
\hbar
\left(
1-\frac{1}{n_\mu}
\right) \, .
\label{matched-heff}
\end{equation}
After removal of the spatial regulator, the corresponding functional
relation is
\begin{equation}
\left[
\hat\phi(\vec{x}),
\hat\pi(\vec{y})
\right]_{\rm eff}
=
i\hbar_{\rm eff}(\mu)
\delta(\vec{x}-\vec{y})
\, .
\label{matched-functional-CCR}
\end{equation}
The matching has a direct interpretation. The local finite-update
construction yields the operator-valued deformation
$\hat C_\epsilon$, whereas coarse-graining evaluates this operator on
the normalized macroscopic representative of the equivalence class.
For the minimal reversible boundary structure, the resulting matrix
element is the fraction of elementary updates that remain internal,
\begin{equation}
\frac{n_\mu-1}{n_\mu}
=
1-\frac{1}{n_\mu} \, .
\end{equation}
The same factor that measures the unresolved internal multiplicity in
the coarse-grained counting therefore determines the effective
canonical commutator.

For $n_\mu=1$, no translated configuration remains within the class,
and Eq.~\eqref{C-expectation-minimal} gives
$\langle\hat C_\epsilon\rangle_\mu=0$. Hence
$\hbar_{\rm eff}=0$ and the effective commutator vanishes. Conversely,
for $n_\mu\to\infty$, the relative contribution of the single boundary
configuration tends to zero,
\begin{equation}
\frac{1}{n_\mu}
\longrightarrow
0 \, ,
\end{equation}
so that
$\langle\hat C_\epsilon\rangle_\mu\to1$,
$\hbar_{\rm eff}\to\hbar$, and the standard canonical algebra is
recovered.

Let us comment that a scale-dependent or running effective Planck constant $\hbar_{\text{eff}}$
draws conceptual parallels with deformed relativity frameworks and condensed-matter
gravity analogs, where fundamental parameters run with the observation scale or
ambient medium dynamics \cite{MagueijoSmolin2003, Volovik2003}. While schemes like
Double Special Relativity enforce an invariant Planck scale through modified dispersion
relations, our framework yields $\hbar_{\text{eff}}(n_\mu)$ as an emergent coarse-graining
parameter directly tied to the unresolvable state updates.

\section{Canonical anticommutation relations from Grassmann variables}
\label{sec:finite-CAR}

We now extend the deterministic finite-update construction to the
degrees of freedom whose infrared representation defines an emergent
fermionic field. This extension realizes explicitly, at the level of
the canonical algebra, a possibility that was left open in
Ref.~\cite{Maiezza:2026wke}. In that work, the microscopic
configuration was generalized schematically to
\begin{equation}
\Phi_{\rm tot}(e_n)
=
\left(
\phi_I(e_n),
A_\mu^a(e_n),
\psi_\alpha(e_n),
\ldots
\right) \, ,
\label{total-microscopic-configuration}
\end{equation}
with the understanding that fermionic statistics and the corresponding
field-theoretic structures should emerge only after coarse-graining.
The symbols in Eq.~\eqref{total-microscopic-configuration} label
microscopic variables according to their eventual infrared
interpretation. In particular, $\psi_\alpha(e_n)$ is not, at this
level, a fundamental Grassmann coordinate or a quantum spinor.

The fundamental dynamics remains the same deterministic,
information-preserving map introduced in
Section~\ref{sec:previous-framework},
\begin{equation}
\Phi_{{\rm tot},n+1}
=
F\left(\Phi_{{\rm tot},n}\right) \, ,
\qquad
\tau_{n+1}-\tau_n
=
\tau_{\min} \, .
\label{universal-update-fermionic-section}
\end{equation}
Thus, the microscopic variables whose infrared images are represented
as fermionic fields are updated, together with all other microscopic
variables, once during every fundamental proper-time interval. No
independent fermionic automaton, additional microscopic evolution law,
or fermion-specific coarse-graining map is introduced.

The discreteness of the microscopic evolution precedes the distinction
between bosonic and fermionic variables. Every component of the
complete microscopic configuration evolves through the same finite
update $F$, independently of the grading eventually assigned to its
infrared representation. What depends on the emergent grading is not
the fundamental dynamics, but the algebraic realization through which
that dynamics is represented after coarse-graining.

In general, the universal map $F$ need not factorize into mutually
independent bosonic and fermionic maps. Its action on a microscopic
variable whose infrared image is fermionic may depend on the complete
microscopic configuration, including variables associated with other
emergent field species. The notation used below to isolate the
fermionic sector therefore denotes the effective representation
induced by the universal map, rather than an autonomous microscopic
dynamics.

Generalizing the logic of Ref.~\cite{Maiezza:2026wke}, and in analogy with
the operator formulation of deterministic models discussed by
't Hooft~\cite{thooft2016cellular}, let $\hat{\mathcal U}$ denote the operator
induced by the universal update in the linear representation of
the deterministic state space,
\begin{equation}
\hat{\mathcal U}\lvert\Phi_{\rm tot}\rangle
=
\left|F\left(\Phi_{\rm tot}\right)\right\rangle \, .
\label{universal-linear-update}
\end{equation}
Since $F$ is bijective, the inverse update exists and is represented by
\begin{equation}
\hat{\mathcal U}^{-1}\lvert\Phi_{\rm tot}\rangle
=
\left|F^{-1}\left(\Phi_{\rm tot}\right)\right\rangle \, .
\label{universal-linear-inverse-update}
\end{equation}
The symmetric part of the elementary forward and backward evolution is
therefore
\begin{equation}
\hat{\mathcal C}
=
\frac{\hat{\mathcal U}+\hat{\mathcal U}^{-1}}{2}
\, .
\label{universal-symmetric-update}
\end{equation}
This is the universal counterpart of the symmetric finite-translation
operator introduced in the bosonic construction. It is defined before
selecting any bosonic or fermionic effective representation and before
projecting onto a coarse-grained equivalence class.

The logical order is essential. First, the finite and reversible
microscopic evolution determines the universal even operator
$\hat{\mathcal C}$. Second, this operator is represented in the algebra
appropriate to the effective infrared variable. In the bosonic sector,
the representation is realized locally through finite translations of
commuting field coordinates. In the fermionic sector, the same update
information is represented within an effective graded algebra. Only
after the corresponding operator-valued canonical relations have been
obtained is the universal microscopic operator projected onto a
coarse-grained equivalence class.

The linear space carrying the coarse-graining representation and the
effective Grassmann algebra are distinct. The complete algebraic
construction is therefore defined on
\begin{equation}
\mathcal A_{\rm cg}\otimes\mathcal A_F \, ,
\label{cg-f-spaces}
\end{equation}
where $\mathcal A_{\rm cg}$ represents the coarse-grained macroscopic
sector and $\mathcal A_F$ denotes the effective Grassmann algebra.
Denoting by $\hat{\mathcal C}_{\rm cg}$ the representation of the
universal symmetric update on the coarse-grained factor, its minimal
action on the complete effective fermionic space is
\begin{equation}
\hat C^{(F)}
\equiv
\hat{\mathcal C}_{\rm cg}\otimes\mathbf 1_F \, ,
\label{fermionic-induced-C}
\end{equation}
where $\mathbf 1_F$ is the identity on the Grassmann algebra.
Equation~\eqref{fermionic-induced-C} does not define a new fundamental
map. It transports the symmetric action of the universal update and
its inverse to the effective fermionic representation.

\begin{remark}[On the minimal factorized representation]
\label{rem:tensor-ansatz}
The identification
$\hat C^{(F)}=\hat{\mathcal C}_{\rm cg}\otimes\mathbf 1_F$
is a working hypothesis rather than a consequence of the Grassmann
algebra. It assumes that the coarse-graining map $C_\mu$, which groups
complete microscopic configurations $\Phi_{\rm tot}$ into equivalence
classes, is insensitive to the infrared grading eventually assigned
to a microscopic label. The Grassmann grading is introduced only in
the effective algebra used to represent the coarse-grained variable,
not as an additional structure acting on the microscopic
configuration space. Coarse-graining and the assignment of the graded
representation therefore act on distinct factors and commute by
construction. This is the minimal factorized representation compatible
with a single deterministic map $F$ and with no independent
fermion-specific microscopic update.
\end{remark}

The fermionic construction is therefore parallel to the bosonic one
at the dynamical level, although the two algebraic representations are
different. For bosonic variables, a local coordinate representation of
the elementary update gives a pair of finite translations in a
commutative field-configuration space. Their antisymmetric and
symmetric combinations define, respectively, the finite-difference
operator and the averaging operator $\hat C^{(B)}_\epsilon$. For
fermionic variables, the same universal update is represented on an
effective graded algebra. The Grassmann coordinates and Berezin
derivatives determine the anticommuting canonical structure, while
$\hat C^{(F)}$ carries the finite-update information inherited from
the microscopic dynamics.

Our objective is consequently the same as in the bosonic sector. We
first construct the minimal effective graded representation of the
operator-valued deformation associated with the finite update. We
then project the resulting operator onto the same coarse-grained
microscopic sector used for the bosonic variables and show that its
scalar matrix element reproduces the effective quantization factor
previously obtained from microscopic counting.

\subsection{One-dimensional Grassmann representation}
\label{subsec:Grassmann-prototype}

We begin with a single Grassmann coordinate $\theta$, satisfying
\begin{equation}
\theta^2=0 \, .
\label{Grassmann-nilpotency}
\end{equation}
The coordinate $\theta$ belongs to the effective graded
representation. It should not be interpreted as a fundamental
anticommuting value carried by the deterministic automaton. The
microscopic variables updated by $F$ are ordinary labels of the
deterministic state space; their Grassmann representation emerges only
in the coarse-grained fermionic description.

On the space in Eq.~\eqref{cg-f-spaces}, the Grassmann coordinate and
its left derivative are represented, in parallel with
Eq.~\eqref{fermionic-induced-C}, as
\begin{equation}
\hat\theta
=
\mathbf 1_{\rm cg}\otimes\theta \, ,
\qquad
\hat\partial_\theta
=
\mathbf 1_{\rm cg}\otimes
\frac{\partial}{\partial\theta} \, ,
\label{Grassmann-effective-operators}
\end{equation}
where $\mathbf 1_{\rm cg}$ is the identity on the coarse-grained
factor. To simplify notation, identity operators and hats on $\theta$
and $\partial/\partial\theta$ will be suppressed whenever no ambiguity
can arise.

An arbitrary function of $\theta$ can be written as
\begin{equation}
\mathcal G(\theta)
=
G_0+\theta G_1 \, .
\label{Grassmann-function}
\end{equation}
Using the left Berezin derivative,
\begin{equation}
\frac{\partial}{\partial\theta}\mathcal G(\theta)
=
G_1 \, .
\label{Berezin-derivative}
\end{equation}
For homogeneous Grassmann functions $\mathcal A$ and $\mathcal B$,
\begin{equation}
\frac{\partial}{\partial\theta}
(\mathcal A\mathcal B)
=
\left(
\frac{\partial\mathcal A}{\partial\theta}
\right)\mathcal B
+
(-1)^{|\mathcal A|}
\mathcal A
\left(
\frac{\partial\mathcal B}{\partial\theta}
\right) \, ,
\label{graded-Leibniz}
\end{equation}
where $|\mathcal A|=0,1$ is the Grassmann parity. In particular,
\begin{equation}
\frac{\partial}{\partial\theta}
(\theta\mathcal G)
=
\mathcal G
-
\theta
\frac{\partial\mathcal G}{\partial\theta} \, .
\label{Grassmann-coordinate-product}
\end{equation}
It follows that
\begin{equation}
\left\{
\theta,
\frac{\partial}{\partial\theta}
\right\}
=
1
\, .
\label{elementary-CAR}
\end{equation}
Thus, the elementary Grassmann anticommutator follows directly from
the graded Leibniz rule. By itself, however, the ordinary Berezin
derivative contains no information about the finite microscopic
update. It characterizes only the graded coordinate structure of the
effective infrared representation.

The finite-update information is instead carried by $\hat C^{(F)}$.
Since $\hat C^{(F)}$ acts on the coarse-grained factor, whereas
$\theta$ and $\partial/\partial\theta$ act on the Grassmann factor,
\begin{equation}
[\hat C^{(F)},\theta]
=
0 \, ,
\qquad
\left[
\hat C^{(F)},
\frac{\partial}{\partial\theta}
\right]
=
0 \, .
\label{CF-commutation}
\end{equation}
These relations follow from the factorized representation
\eqref{cg-f-spaces} and introduce no additional hypothesis beyond
Remark~\ref{rem:tensor-ansatz}.

At the microscopic level, the variable whose infrared image is
represented by $\theta$ evolves only through the finite map $F$.
Therefore, its effective Grassmann representation should not be
assigned an independent microscopic evolution law. The Berezin
derivative encodes the graded differential structure of the infrared
representation, whereas $\hat C^{(F)}$ encodes the symmetric
finite-update structure inherited from the microscopic dynamics. The
minimal effective representation must combine these two ingredients.

Within the factorized construction adopted above, the minimal
update-dressed Berezin operator is
\begin{equation}
\mathfrak D_\theta
=
\hat C^{(F)}
\frac{\partial}{\partial\theta}
\, .
\label{deformed-Grassmann-derivative}
\end{equation}
This choice preserves the odd parity of the Berezin derivative because
$\hat C^{(F)}$ is even, reduces to the ordinary Berezin derivative
when $\hat C^{(F)}$ acts as the identity, and introduces no independent
fermionic update operator. The terminology ``update-dressed''
emphasizes that the Berezin derivative belongs to the effective
Grassmann representation, whereas the even operator multiplying it is
inherited from the universal discrete update.

Equation~\eqref{deformed-Grassmann-derivative} is not a finite
difference of the Grassmann coordinate analogous to the bosonic
central difference. The Grassmann coordinate is not a fundamental
discrete coordinate. Rather, the equation specifies the minimal
factorized representation of the same symmetric microscopic update
within the effective graded algebra.

Using the tensor-product representation, one has
\begin{align}
\{\theta,\mathfrak D_\theta\}
&=
\left\{
\mathbf 1_{\rm cg}\otimes\theta,
\hat{\mathcal C}_{\rm cg}\otimes
\frac{\partial}{\partial\theta}
\right\}
\nonumber
\\
&=
\hat{\mathcal C}_{\rm cg}\otimes
\left\{
\theta,
\frac{\partial}{\partial\theta}
\right\}
\nonumber
\\
&=
\hat{\mathcal C}_{\rm cg}\otimes\mathbf 1_F
=
\hat C^{(F)} \, .
\label{deformed-Grassmann-CAR-derivation}
\end{align}
Therefore,
\begin{equation}
\{\theta,\mathfrak D_\theta\}
=
\hat C^{(F)}
\, .
\label{deformed-Grassmann-CAR}
\end{equation}
Within the minimal representation adopted above, the operator-valued
right-hand side is inherited from the universal finite update on the
complete microscopic configuration, while the anticommutator follows
from the graded Leibniz rule. The universal symmetric update
determines the even operator $\hat C^{(F)}$, whereas the Grassmann
algebra determines that this operator appears in an anticommutator
rather than in a commutator. The scalar coarse-graining factor
$Q(n_\mu)$ has not entered the construction at this stage.

Equation~\eqref{deformed-Grassmann-CAR} is the fermionic counterpart
of the operator-valued bosonic relation
\begin{equation}
[\hat x,\hat p_\epsilon]
=
i\hbar\hat C_\epsilon \, .
\label{bosonic-prototype-recalled}
\end{equation}
The two operators have the same dynamical origin in the universal
discrete update but different algebraic realizations. This parallelism
does not imply an identity between an ordinary finite difference and a
Berezin derivative. The universal object is the finite microscopic
update; its effective differential representation depends on the
grading of the emergent coordinate.

Introducing the canonically normalized effective conjugate operator
\begin{equation}
\hat\pi_\theta
=
i\hbar\mathfrak D_\theta \, ,
\label{Grassmann-momentum}
\end{equation}
one obtains
\begin{equation}
\{\theta,\hat\pi_\theta\}
=
i\hbar\hat C^{(F)}
\, .
\label{deformed-Grassmann-momentum-CAR}
\end{equation}
This is a kinematical definition at the level of the effective
canonical algebra. Its relation to the momentum derived from a
specific infrared fermionic Lagrangian depends on the effective action
and on the adopted left- or right-derivative convention.

The logical order is therefore parallel to the bosonic construction.
The universal finite update is specified first, and its symmetric
part is represented in the minimal effective graded algebra. The
resulting operator-valued relation is then obtained exactly within
that representation. Only afterwards is the operator projected onto a
coarse-grained equivalence class. The scalar factor $Q(n_\mu)$ is not
used to define the fermionic algebra.

\subsection{Fermionic representation on a spatial lattice}
\label{subsec:fermionic-lattice}

We now extend the graded representation to a fermionic field on a
spatial lattice. The spatial lattice is introduced only as a regulator
of the effective field theory. It should not be confused with the
network of fundamental events or with the discrete proper-time
evolution generated by $F$.

Let $\theta_{\alpha,j}$ denote the Grassmann coordinate associated
with lattice site $j$ and spinorial or internal index $\alpha$. The
coordinates and their left derivatives satisfy
\begin{align}
\{\theta_{\alpha,j},\theta_{\beta,k}\}
&=
0 \, ,
\nonumber
\\
\left\{
\frac{\partial}{\partial\theta_{\alpha,j}},
\frac{\partial}{\partial\theta_{\beta,k}}
\right\}
&=
0 \, ,
\nonumber
\\
\left\{
\theta_{\alpha,j},
\frac{\partial}{\partial\theta_{\beta,k}}
\right\}
&=
\delta_{\alpha\beta}\delta_{jk} \, .
\label{lattice-Grassmann-algebra}
\end{align}
In the minimal factorized construction, every effective fermionic
direction inherits the same symmetric update of the complete
microscopic configuration. The indices $(\alpha,j)$ label the
Grassmann coordinate on which the Berezin derivative acts; they do not
label independent microscopic update maps or independent
coarse-graining procedures.

Accordingly,
\begin{equation}
\theta_{\alpha,j}
\equiv
\mathbf 1_{\rm cg}\otimes\theta_{\alpha,j} \, ,
\qquad
\hat C^{(F)}
=
\hat{\mathcal C}_{\rm cg}\otimes\mathbf 1_F \, .
\label{lattice-factorized-operators}
\end{equation}
It follows that
\begin{equation}
[\hat C^{(F)},\theta_{\alpha,j}]
=
0 \, ,
\qquad
\left[
\hat C^{(F)},
\frac{\partial}{\partial\theta_{\alpha,j}}
\right]
=
0 \, .
\label{lattice-CF-commutation}
\end{equation}
For each Grassmann direction, define
\begin{equation}
\mathfrak D_{\alpha,j}
=
\hat C^{(F)}
\frac{\partial}{\partial\theta_{\alpha,j}} \, .
\label{lattice-deformed-Grassmann-derivative}
\end{equation}
Then
\begin{equation}
\{\theta_{\alpha,j},\mathfrak D_{\beta,k}\}
=
\delta_{\alpha\beta}\delta_{jk}
\hat C^{(F)}
\, .
\label{lattice-deformed-CAR}
\end{equation}
Defining
\begin{equation}
\hat\pi_{\theta,\beta,k}
=
i\hbar\mathfrak D_{\beta,k} \, ,
\label{lattice-fermionic-momentum}
\end{equation}
we obtain
\begin{equation}
\{\theta_{\alpha,j},\hat\pi_{\theta,\beta,k}\}
=
i\hbar
\delta_{\alpha\beta}\delta_{jk}
\hat C^{(F)}
\, .
\label{lattice-deformed-momentum-CAR}
\end{equation}
This relation is directly parallel to the bosonic result
\begin{equation}
[\hat\phi_j,\hat\pi_{\epsilon,k}]
=
i\hbar\delta_{jk}\hat C_{\epsilon,k} \, .
\label{bosonic-relation-for-comparison}
\end{equation}
The commutator and anticommutator reflect the different gradings of the
bosonic and fermionic effective variables. The two operators encode
different algebraic representations of the same finite and reversible
microscopic evolution.

A more general effective representation could associate
direction-dependent operators $\hat C^{(F)}_{\alpha,j}$ with different
fermionic directions. Such a construction would require additional
microscopic information and lies beyond the minimal factorized
realization considered here.

\subsection{Projection onto the coarse-grained microscopic sector}
\label{subsec:fermionic-matching}

We now connect the operator-valued deformation in
Eq.~\eqref{lattice-deformed-momentum-CAR} with the effective
quantization factor previously obtained from microscopic counting.
This is the fermionic counterpart of the matching performed in the
bosonic sector.

The fundamental equivalence classes are classes of complete
microscopic configurations, rather than independent classes assigned
to each emergent field species. Let
\begin{equation}
[\Phi_\mu]
=
\left\{
\Phi_{\rm tot}\in\Lambda
\,\middle|\,
C_\mu(\Phi_{\rm tot})=\Phi_\mu
\right\} \, ,
\qquad
n_\mu
=
|[\Phi_\mu]| \, ,
\label{universal-equivalence-class-fermionic}
\end{equation}
where $C_\mu$ is the single coarse-graining map at resolution $\mu$.
The same equivalence class contains all microscopic data whose
infrared representations may be bosonic, fermionic, or of any other
effective type. No separate fermionic equivalence class is introduced.

Following the same uniform linear representation used in the bosonic
matching, the normalized macroscopic representative is
\begin{equation}
|\Phi_\mu\rangle
=
\frac{1}{\sqrt{n_\mu}}
\sum_{\Phi_{\rm tot}\in[\Phi_\mu]}
|\Phi_{\rm tot}\rangle
\, .
\label{universal-coarse-state-fermionic}
\end{equation}
This state is the linear representation of a macroscopic equivalence
class and should not be interpreted as a fundamental quantum
superposition. The Grassmann algebra provides the graded
representation, but it does not carry an independent coarse-graining
operation. The projection acts on the single linear space of complete
microscopic configurations.

In complete analogy with
Subsec.~\ref{subsec:finite-matching}, one has
\begin{equation}
\langle\Phi_\mu|
\hat{\mathcal C}
|\Phi_\mu\rangle
=
1-\frac{b_\mu}{n_\mu}
\, .
\label{universal-C-expectation-reversible-fermionic}
\end{equation}
Here $b_\mu$ is the common number of forward and backward outgoing
transitions. Its equality in the two directions follows from the
reversibility of the universal update, as in the bosonic construction.

We now specialize to the same minimal boundary-crossing model used in
the bosonic sector and in Ref.~\cite{Maiezza:2026wke},
\begin{equation}
b_\mu
=
1 \, .
\label{minimal-fermionic-boundary-crossing}
\end{equation}
The equality of the forward and backward crossing numbers follows from
reversibility, whereas $b_\mu=1$ is the minimal counting assumption.
It follows that
\begin{equation}
\langle\Phi_\mu|
\hat{\mathcal C}
|\Phi_\mu\rangle
=
1-\frac{1}{n_\mu}
=
Q(n_\mu)
\, .
\label{fermionic-C-expectation}
\end{equation}
This is the same matrix element obtained in the bosonic matching
because the operator being projected, the equivalence class, and the
coarse-graining map are the same. The two representations differ only
in the canonical algebra through which the universal update is
realized.

On the complete effective fermionic space, define
\begin{equation}
\hat{\boldsymbol{\mathcal P}}_\mu
=
\hat{\mathcal P}_\mu\otimes\mathbf 1_F \, ,
\qquad
\hat{\mathcal P}_\mu
=
|\Phi_\mu\rangle\langle\Phi_\mu| \, .
\label{fermionic-macroscopic-projector}
\end{equation}
Then
\begin{align}
\hat{\boldsymbol{\mathcal P}}_\mu
\hat C^{(F)}
\hat{\boldsymbol{\mathcal P}}_\mu
&=
\left(
\hat{\mathcal P}_\mu
\hat{\mathcal C}_{\rm cg}
\hat{\mathcal P}_\mu
\right)
\otimes\mathbf 1_F
\nonumber
\\
&=
Q(n_\mu)
\hat{\boldsymbol{\mathcal P}}_\mu \, .
\label{projected-fermionic-C-derivation}
\end{align}
Therefore,
\begin{equation}
\hat{\boldsymbol{\mathcal P}}_\mu
\hat C^{(F)}
\hat{\boldsymbol{\mathcal P}}_\mu
=
Q(n_\mu)
\hat{\boldsymbol{\mathcal P}}_\mu
\, .
\label{projected-fermionic-C}
\end{equation}
This equality does not result from a second fermionic counting
argument. It is inherited from the projection of the same universal
update onto the same equivalence class of complete microscopic
configurations.

Applying Eq.~\eqref{projected-fermionic-C} to
Eq.~\eqref{lattice-deformed-momentum-CAR} gives
\begin{align}
\hat{\boldsymbol{\mathcal P}}_\mu
\{\theta_{\alpha,j},\hat\pi_{\theta,\beta,k}\}
\hat{\boldsymbol{\mathcal P}}_\mu
=
i\hbar
Q(n_\mu)
\delta_{\alpha\beta}\delta_{jk}
\hat{\boldsymbol{\mathcal P}}_\mu \, .
\label{projected-fermionic-CAR}
\end{align}
Equivalently, the restricted matrix elements in the macroscopic sector
are reproduced by the effective lattice relation
\begin{equation}
\{\theta_{\alpha,j},\hat\pi_{\theta,\beta,k}\}_{\rm eff}
=
i\hbar_{\rm eff}(\mu)
\delta_{\alpha\beta}\delta_{jk}
\, ,
\label{effective-fermionic-CAR}
\end{equation}
with
\begin{equation}
\hbar_{\rm eff}(\mu)
=
\hbar Q(n_\mu)
=
\hbar
\left(
1-\frac{1}{n_\mu}
\right)
\, .
\label{fermionic-heff}
\end{equation}
The matching has the same logical form as in the bosonic sector. The
universal finite update is first represented through an
operator-valued canonical relation in the effective graded algebra.
The common coarse-graining projection then replaces the symmetric
update operator by its scalar matrix element $Q(n_\mu)$. The effective
Planck constant obtained in Ref.~\cite{Maiezza:2026wke} is thus
recovered as the coarse-grained normalization of the fermionic
finite-update representation.

\subsection{Universal effective quantization scale}
\label{subsec:universal-heff}

Within the common microscopic update and the minimal factorized
effective representation adopted above, the bosonic and fermionic
variables arise from the same complete microscopic configuration,
evolve through the same deterministic map $F$, and are observed
through the same coarse-graining map $C_\mu$. Their canonical
structures are therefore not normalized by independently assigned
microscopic cardinalities. The relevant quantity is
\begin{equation}
n_\mu
=
|[\Phi_\mu]| \, .
\label{universal-cardinality}
\end{equation}
Before specializing to the minimal model, the common projected
normalization is
\begin{equation}
Q_\mu
=
1-\frac{b_\mu}{n_\mu} \, .
\label{general-universal-Q}
\end{equation}
It is a property of the projected microscopic update and does not
depend on whether an infrared variable is represented by a commuting
or Grassmann coordinate. In the minimal reversible realization,
\begin{equation}
b_\mu
=
1 \, ,
\qquad
Q_\mu
=
Q(n_\mu)
=
1-\frac{1}{n_\mu} \, .
\label{minimal-universal-Q}
\end{equation}
The bosonic construction gives
\begin{equation}
[\hat\phi_j,\hat\pi_{\epsilon,k}]
=
i\hbar\delta_{jk}
\hat C_{\epsilon,k} \, ,
\label{bosonic-operator-valued-relation-recalled}
\end{equation}
and
\begin{equation}
\hat{\mathcal P}_\mu
\hat C^{(B)}_{\epsilon,k}
\hat{\mathcal P}_\mu
=
Q(n_\mu)
\hat{\mathcal P}_\mu \, .
\label{bosonic-Q-recalled}
\end{equation}
The fermionic construction gives
\begin{equation}
\{\theta_{\alpha,j},\hat\pi_{\theta,\beta,k}\}
=
i\hbar
\delta_{\alpha\beta}\delta_{jk}
\hat C^{(F)} \, ,
\label{fermionic-operator-valued-relation-recalled}
\end{equation}
and
\begin{equation}
\hat{\boldsymbol{\mathcal P}}_\mu
\hat C^{(F)}
\hat{\boldsymbol{\mathcal P}}_\mu
=
Q(n_\mu)
\hat{\boldsymbol{\mathcal P}}_\mu \, .
\label{fermionic-Q-recalled}
\end{equation}
Thus, the two canonical sectors do not possess independent
coarse-graining factors that subsequently happen to coincide. They
inherit the same scalar matrix element of one projected universal
update,
\begin{equation}
Q(n_\mu)
=
1-\frac{1}{n_\mu}
\, .
\label{universal-Q}
\end{equation}
If sector labels are introduced, one may write
\begin{equation}
Q_B(n_\mu)
\equiv
Q(n_\mu) \, ,
\qquad
Q_F(n_\mu)
\equiv
Q(n_\mu) \, ,
\label{universal-Q-identification}
\end{equation}
but $Q_B$ and $Q_F$ are not independently defined microscopic
quantities. They denote the appearance of the same normalization in
two differently graded canonical representations.

Accordingly,
\begin{equation}
\hbar_{\rm eff}^{(B)}(\mu)
=
\hbar Q(n_\mu) \, ,
\qquad
\hbar_{\rm eff}^{(F)}(\mu)
=
\hbar Q(n_\mu) \, ,
\label{sector-effective-Planck-constants}
\end{equation}
and therefore
\begin{equation}
\hbar_{\rm eff}^{(B)}(\mu)
=
\hbar_{\rm eff}^{(F)}(\mu)
=
\hbar_{\rm eff}(\mu)
=
\hbar
\left(
1-\frac{1}{n_\mu}
\right)
\, .
\label{universal-heff}
\end{equation}
The equality follows because the two canonical algebras represent the
same finite microscopic update and are reduced by the same
coarse-graining map on the same equivalence class. The canonical
algebras remain different, while their common normalization reflects
the same unresolved microscopic multiplicity.

At proper-time and spatial scales sufficiently larger than the
fundamental cutoff,
\begin{equation}
\theta_{\alpha,j}
\longrightarrow
\psi_\alpha(\vec{x}) \, ,
\qquad
\delta_{jk}
\longrightarrow
\delta(\vec{x}-\vec{y}) \, ,
\label{fermionic-continuum-correspondence}
\end{equation}
where the second replacement includes the appropriate lattice
normalization. The resulting effective functional relation is
\begin{equation}
\{
\hat\psi_\alpha(\vec{x}),
\hat\pi_{\psi,\beta}(\vec{y})
\}_{\rm eff}
=
i\hbar_{\rm eff}(\mu)
\delta_{\alpha\beta}
\delta(\vec{x}-\vec{y})
\, .
\label{continuum-effective-CAR}
\end{equation}
This relation is written in canonical coordinate-momentum form. Its
translation into a field-adjoint anticommutator depends on the
specific effective fermionic action selected in the infrared. It is an
effective macroscopic relation. At the fundamental level, evolution
remains a sequence of deterministic updates separated by
$\tau_{\min}$.

For $n_\mu=1$,
\begin{equation}
Q(1)
=
0 \, ,
\label{universal-deterministic-Q}
\end{equation}
and
\begin{equation}
[\hat\phi_j,\hat\pi_k]_{\rm eff}
=
0 \, ,
\qquad
\{
\theta_{\alpha,j},
\hat\pi_{\theta,\beta,k}
\}_{\rm eff}
=
0 \, .
\label{universal-deterministic-algebras}
\end{equation}
This is the deterministic limit. It should not be interpreted as a
fundamental continuum QFT with $\hbar=0$; the appropriate description
is the deterministic update generated by $F$.

In the opposite limit,
\begin{equation}
n_\mu
\longrightarrow
\infty
\qquad\Longrightarrow\qquad
Q(n_\mu)
\longrightarrow
1 \, ,
\label{universal-infrared-Q}
\end{equation}
the standard bosonic and fermionic canonical relations are recovered
simultaneously. The resulting effective algebra is
\begin{align}
[\hat\phi_j,\hat\pi_k]_{\rm eff}
&=
i\hbar Q(n_\mu)\delta_{jk} \, ,
\nonumber
\\[1mm]
\{
\theta_{\alpha,j},
\hat\pi_{\theta,\beta,k}
\}_{\rm eff}
&=
i\hbar Q(n_\mu)
\delta_{\alpha\beta}\delta_{jk} \, ,
\nonumber
\\[1mm]
Q(n_\mu)
&=
1-\frac{1}{n_\mu} \, .
\label{unified-canonical-algebra}
\end{align}
Thus, the commutator and anticommutator express the different gradings
of the bosonic and fermionic representations, while their common
normalization is inherited from the same universal deterministic
update and the same loss of microscopic resolution under
coarse-graining.

This result is consistent with the statistical argument of
Ref.~\cite{Maiezza:2026wke}, although an explicit fermionic sector was
not constructed there. The two approaches are complementary: the
statistical argument identifies the microscopic origin of the factor
$Q(n_\mu)$, while the finite-update construction shows how the same
factor enters differently graded canonical algebras. The effective
quantization scale therefore need not be introduced independently for
each field species, since it is inherited from the common microscopic
update and coarse-graining map.

\subsection{Synoptic picture of the emergent QFT}
\label{subsec:picture}

The successive levels of the construction, initiated in
Ref.~\cite{Maiezza:2026wrp,Maiezza:2026wke} and developed here, are summarized in
Table~\ref{tab:unified-framework}. The table may be read in two
complementary directions. From the microscopic scale toward the
infrared, it describes how deterministic updates, coarse-graining,
canonical quantization, QFT, and a smooth relativistic geometry arise
in successive effective representations. From the infrared toward
shorter proper-time scales, it describes the progressive loss of
validity of the continuum quantum and geometric descriptions.

\begingroup
\renewcommand{\arraystretch}{1.25}
\setlength{\tabcolsep}{4pt}

\begin{xltabular}{\textwidth}{
@{}
>{\raggedright\arraybackslash}p{0.15\textwidth}
>{\raggedright\arraybackslash}p{0.23\textwidth}
>{\raggedright\arraybackslash}X
>{\raggedright\arraybackslash}p{0.20\textwidth}
@{}
}

\caption{
Logical structure of the minimal proper-time framework, from the
deterministic microscopic dynamics to the infrared quantum and
geometric descriptions.
}
\label{tab:unified-framework}
\\

\toprule
\textbf{Level}
&
\textbf{Mathematical description}
&
\textbf{Physical meaning}
&
\textbf{Emergent structure}
\\
\midrule
\endfirsthead

\multicolumn{4}{l}{
\small\itshape
Table~\ref{tab:unified-framework} continued from the previous page
}
\\[1mm]
\toprule
\textbf{Level}
&
\textbf{Mathematical description}
&
\textbf{Physical meaning}
&
\textbf{Emergent structure}
\\
\midrule
\endhead

\midrule
\multicolumn{4}{r}{
\small\itshape Continued on the next page
}
\\
\endfoot

\bottomrule
\endlastfoot

Microscopic dynamics
&
\(
\tau_n=n\tau_{\min}
\),
\(
\Phi_{{\rm tot},n+1}
=
F(\Phi_{{\rm tot},n})
\)
&
During each elementary proper-time interval, the complete microscopic
configuration is updated by a deterministic, bijective, and
information-preserving map. The corresponding linear operator
represents this permutation dynamics without introducing fundamental
quantum behavior.
&
Discrete evolution governed by $F$
\\
\addlinespace

Coarse-graining
&
\(
[\Phi_\mu]
=
\{
\Phi_{\rm tot}
\mid
C_\mu(\Phi_{\rm tot})=\Phi_\mu
\}
\),
\(
n_\mu=|[\Phi_\mu]|
\)
&
Microscopic configurations that cannot be distinguished at resolution
$\mu$ are represented by the same macroscopic state. Boundary
transitions between equivalence classes make the reduced description
effectively open.
&
Unresolved multiplicity and effective non-unitarity
\\
\addlinespace

Effective quantization
&
\(
Q(n_\mu)
=
1-\dfrac{1}{n_\mu}
\),
\(
\hbar_{\rm eff}
=
\hbar Q(n_\mu)
\)
&
The unresolved internal fraction determines the strength of the
effective quantum description. Quantization disappears for
$n_\mu=1$ and approaches its standard normalization for
$n_\mu\gg1$.
&
Running effective Planck constant
\\
\addlinespace

Canonical representation
&
\(
[\hat\phi_j,\hat\pi_k]_{\rm eff}
=
i\hbar_{\rm eff}\delta_{jk}
\),
\(
\{\theta_{\alpha,j},\hat\pi_{\theta,\beta,k}\}_{\rm eff}
=
i\hbar_{\rm eff}\delta_{\alpha\beta}\delta_{jk}
\)
&
Bosonic and fermionic variables give differently graded
representations of the same finite microscopic evolution. Their
canonical normalizations coincide because both inherit the same
coarse-graining factor within the minimal factorized representation.
&
Common bosonic and fermionic canonical scale
\\
\addlinespace

Extended infrared dynamics
&
\(
i\partial_\tau\Psi
=
\hat H'\Psi
\)
&
When individual microscopic updates are no longer resolved, the
cumulative deterministic evolution admits a smooth
Nambu--Schr\"odinger representation. The relative boundary leakage
becomes increasingly suppressed.
&
Effective proper-time QFT
\\
\addlinespace

Physical quantum sector
&
\(
\Psi_R
=
\displaystyle
\int_{\tau_{\min}}^\infty
d\tau\,
\Psi(\tau)
\)
&
The proper-time evolution is projected onto the physical
four-dimensional sector. In the infrared,
$\hbar_{\rm eff}\to\hbar$ and the standard unitary canonical
description is recovered.
&
Ordinary QFT in the Schr\"odinger representation
\\
\addlinespace

Infrared geometry
&
\(
n_\mu\gg1
\)
and smooth coarse-grained metric data
&
After sufficient coarse-graining, the microscopic event structure
admits a differentiable spacetime representation. Under the infrared
assumptions of the top-down construction, its macroscopic geometric
dynamics is described by general relativity.
&
Emergent relativistic spacetime and GR
\\

\end{xltabular}
\endgroup

\section{Possible signatures of the deterministic crossover}
\label{sec:phenomenological-signatures}

In this section, we briefly discuss two possible phenomenological
manifestations of the deterministic crossover. The first concerns
radiative corrections in high-energy scattering amplitudes. The
second concerns Hawking radiation, whose existence relies on the
quantization of matter fields in a gravitational background \cite{Hawking:1974rv,Hawking:1977wz}.

The discussion is exploratory. Its purpose is to identify physical
questions that follow from the deterministic crossover rather than to
provide complete phenomenological predictions, beyond the scope of
this work.

\subsection{Tree and loop contributions at high energies}
\label{subsec:tree-loop-signature}

Consider schematically a scattering amplitude written as a
perturbative expansion,
\begin{equation}
\mathcal A
=
\mathcal A_{\rm tree}
+
\mathcal A_{\rm 1-loop}
+
\mathcal A_{\rm 2-loop}
+
\cdots \, .
\label{amplitude-loop-expansion}
\end{equation}
The tree-level contribution describes the leading interaction allowed
by the classical field equations, whereas loop contributions encode
quantum fluctuations. Since the effective quantization scale decreases
toward the microscopic deterministic regime, the relative importance
of radiative corrections is expected to decrease as well.

This expectation is consistent with the proper-time formulation. The
minimal proper-time scale suppresses propagators at large Euclidean
momentum and consequently damps the ultraviolet region of loop
integrals. At energies well below the fundamental scale, this
suppression is negligible and the usual radiative corrections of QFT
are recovered. At increasingly high energies, loop contributions
probe the proper-time cutoff and become progressively suppressed.

The damping of proper-time propagators and the decrease of
$\hbar_{\rm eff}$ should be regarded as two related effective
descriptions of the same crossover, rather than as independent
suppression mechanisms. A complete calculation is required to
determine how the ultraviolet modification of the propagators is
expressed in terms of the coarse-graining factor $Q(n_\mu)$.

A useful quantity is the relative radiative contribution to a cross
section,
\begin{equation}
R_{\rm rad}(E)
=
\frac{
\sigma_{\rm full}(E)
-
\sigma_{\rm tree}(E)
}{
\sigma_{\rm tree}(E)
} \, .
\label{relative-radiative-correction}
\end{equation}
In standard QFT, $R_{\rm rad}$ is determined by the running couplings,
the available phase space, and the relevant loop functions. In the
present framework, an additional suppression is expected when the
characteristic proper-time scale of the process approaches
$\tau_{\min}$.

For notational convenience, we define
\begin{equation}
Q(E)
\equiv
Q\left(n_{\mu(E)}\right) \, ,
\label{energy-dependent-Q}
\end{equation}
where $\mu(E)$ denotes the coarse-graining scale associated with the
characteristic proper-time resolution of the process. The relation
between the collision energy and this resolution scale is generally
process dependent and is not fixed by the present construction.

Schematically, one may write
\begin{equation}
R_{\rm rad}^{\rm eff}(E)
=
\mathcal S_{\rm rad}
\left[
Q(E)
\right]
R_{\rm rad}^{\rm QFT}(E) \, ,
\label{radiative-suppression-function}
\end{equation}
where
\begin{equation}
\mathcal S_{\rm rad}(1)
=
1,
\qquad
\mathcal S_{\rm rad}(0)
=
0 \, .
\label{radiative-suppression-limits}
\end{equation}
Equation~\eqref{radiative-suppression-function} is solely a schematic
parametrization. The
suppression function may depend on the process, the relevant
kinematic scales, and the perturbative order. Its form, together with
the relation between the collision energy and the corresponding
proper-time resolution, must be derived from the proper-time
amplitudes and should not be fixed by a phenomenological ansatz.

A conceptually transparent test would compare a tree-dominated process
with a loop-induced process at the same characteristic hard scale.
If the deterministic crossover is approached, the loop-induced
channel should be suppressed relative to the tree-level reference.
The relevant signature would therefore not be a universal reduction
of all scattering rates, but a decreasing relative weight of
contributions that exist only because of quantum fluctuations. In a
concrete comparison, differences in couplings, phase space, and
kinematics would have to be treated explicitly.

In the formal deterministic limit, this behavior may be summarized as
\begin{equation}
\frac{
\mathcal A_{\rm loop}
}{
\mathcal A_{\rm tree}
}
\longrightarrow
0
\qquad
\text{as}
\qquad
Q(n_\mu)
\longrightarrow
0 \, .
\label{loop-tree-deterministic-limit}
\end{equation}
Equation~\eqref{loop-tree-deterministic-limit} should be understood as
a limiting physical expectation. It does not imply that the continuum
loop expansion remains valid at the fundamental scale. It describes
instead how the QFT representation should approach its boundary of
validity from the infrared side.

The ultraviolet freezing of the running couplings found in the
proper-time formulation provides an effective QFT representation of
the same transition. It should not be interpreted as evidence that
QFT remains the fundamental description at arbitrarily high energies.
Rather, the suppression of loop running is the remnant, within the
extrapolated QFT language, of the progressive disappearance of
quantum fluctuations. At the microscopic level, the evolution is
governed by the deterministic map $F$.

\subsection{Effective Hawking radiation}
\label{subsec:effective-Hawking-radiation}

A second possible implication concerns Hawking radiation. This
application is especially natural after extending the canonical
construction to both bosonic and fermionic degrees of freedom, since a
black hole can emit fields obeying either statistics.

For a stationary black hole with surface gravity $\kappa$, the
standard Hawking temperature is
\begin{equation}
T_H
=
\frac{
\hbar\kappa
}{
2\pi
} \, ,
\label{standard-Hawking-temperature}
\end{equation}
in units in which $c=k_B=1$. As a first heuristic approximation, let
us assume that the exterior semiclassical geometry remains fixed and
that the modification of field quantization enters the Hawking
derivation only through the replacement
\begin{equation}
\hbar
\longrightarrow
\hbar_{\rm eff}(\mu)
=
\hbar Q(n_\mu) \, .
\label{Hawking-heff-replacement}
\end{equation}
Under these assumptions, one obtains the effective temperature
\begin{equation}
T_H^{\rm eff}
=
Q(n_\mu)\,
T_H
\, .
\label{effective-Hawking-temperature}
\end{equation}
Equation~\eqref{effective-Hawking-temperature} is conditional on the
fixed-background approximation. It has not been derived here from a
first principles, as a modified calculation of tunnelling 
amplitudes \cite{Parikh:1999mf} \footnote{Extending the semiclassical tunnelling picture to account for UV energy cutoffs or minimal length scales similarly leads to a suppression of high-energy emission modes and resolves trans-Planckian pathologies \cite{Hossenfelder2003, Hossenfelder2004}. The factor $Q(n_\mu)$ provides an explicit microscopic rationale for such energy-dependent suppression factors in the probability rate of quantum tunnelling.}.
Moreover, the relevant coarse-graining scale in a black-hole geometry is not yet determined.
In general, it may depend on the emitted mode and on the geometric
parameters, through quantities such as $\omega$, $\kappa$, or the
local curvature.

The standard Hawking result is recovered in the infrared regime,
\begin{equation}
n_\mu
\longrightarrow
\infty
\qquad\Longrightarrow\qquad
T_H^{\rm eff}
\longrightarrow
T_H \, .
\label{Hawking-infrared-limit}
\end{equation}
Conversely, within the same fixed-background approximation, the
effective temperature is suppressed as the deterministic regime is
approached,
\begin{equation}
n_\mu
\longrightarrow
1
\qquad\Longrightarrow\qquad
T_H^{\rm eff}
\longrightarrow
0 \, .
\label{Hawking-deterministic-limit}
\end{equation}
This behavior has a direct qualitative interpretation. Hawking
radiation is intrinsically quantum and should therefore weaken when
the effective quantization scale decreases. Its suppression would be
the gravitational counterpart of the suppression of radiative
corrections in high-energy particle processes.
This suppression mechanism aligns naturally with phenomenological predictions derived
from the Generalized Uncertainty Principle, where deformations of canonical
commutation relations lead to black hole remnants and a complete shutdown of Hawking
radiation as the black hole approaches the Planck scale \cite{Scardigli1999,Adler2001}.
In our framework, the vanishing of $T_H^{\text{eff}}$ arises not from an ad hoc modification
of phase space, but as a direct dynamical consequence of the underlying deterministic
update rate at $n_\mu \to 1$.
The limit $n_\mu\to1$, however, cannot be described completely while keeping
the semiclassical geometry fixed, since the geometric description
itself approaches the boundary of its validity.

Provisionally retaining the standard greybody factors, the
corresponding occupation numbers may be written schematically as
\begin{align}
\left\langle
N_{\omega,s}^{(B)}
\right\rangle_{\rm eff}
&=
\frac{
\Gamma_{\omega,s}^{(B)}
}{
\exp\left(
\omega/T_H^{\rm eff}
\right)
-
1
} \, ,
\label{effective-bosonic-Hawking-spectrum}
\\[1mm]
\left\langle
N_{\omega,s}^{(F)}
\right\rangle_{\rm eff}
&=
\frac{
\Gamma_{\omega,s}^{(F)}
}{
\exp\left(
\omega/T_H^{\rm eff}
\right)
+
1
} \, ,
\label{effective-fermionic-Hawking-spectrum}
\end{align}
where $\Gamma_{\omega,s}^{(B,F)}$ denote the corresponding greybody
factors. The Bose--Einstein and Fermi--Dirac distributions remain
different, as do their greybody factors, but their effective
temperature is governed by the same factor $Q(n_\mu)$. Within the
approximations stated above, this common temperature reflects the
common normalization of the bosonic and fermionic canonical algebras
derived in the present work.

Equations~\eqref{effective-Hawking-temperature}--
\eqref{effective-fermionic-Hawking-spectrum} should be regarded as a
first heuristic estimate. A complete calculation would require the
determination of the relevant coarse-graining scale in the black-hole
geometry, a derivation of the modified emission spectrum, and the
inclusion of possible corrections to the greybody factors and to the
semiclassical background.

Let $\mathcal P_H$ denote the total power emitted into bosonic and
fermionic degrees of freedom. If one approximates the emission
as a four-dimensional thermal flux, keeps the geometry and emitting
area fixed, and neglects the frequency dependence induced through the
greybody factors, the usual thermal scaling suggests
\begin{equation}
\mathcal P_H^{\rm eff}
\sim
Q(n_\mu)^4
\mathcal P_H \, .
\label{heuristic-Hawking-power}
\end{equation}
\textit{A priori}, the complete dependence on $Q$ need not be a pure fourth
power once the approximations are relaxed. Therefore, the power in
Eq.~\eqref{heuristic-Hawking-power} is only a dimensional estimate.

The main qualitative question is whether $Q(n_\mu)$ decreases rapidly
enough during the final evolution to suppress the semiclassical
evaporation before its standard endpoint. Depending on the relation
between the equivalence-class cardinality, the relevant mode scale,
and the black-hole parameters, the framework could describe a slowing
of the evaporation, an asymptotic suppression, or a crossover to the
non-geometric deterministic substrate. The present construction does
not by itself select among these possibilities.
In particular, the limit $T_H^{\rm eff}\to0$ should not automatically
be interpreted as the prediction of a stable remnant. Near
$n_\mu=1$, the smooth metric and the semiclassical field description
approach the boundary of their validity. The physically relevant
statement is therefore that quantum radiation is progressively
suppressed as the deterministic substrate is approached.

\subsection{Phenomenological status, falsifiability, and no-go constraints}
\label{sec:phenomenology-falsifiability}

A recurring question for deterministic approaches to emergent quantum
theory concerns their phenomenological status. The present framework
establishes a model-independent relation between deterministic finite
updates, coarse-graining, and the emergence of bosonic and fermionic
operator algebras. By itself, this algebraic result does not provide a
parameter-free prediction for a specific experiment. Quantitative
predictions require a concrete realization of the microscopic map
$F$, the fundamental proper-time scale $\tau_{\min}$, and the embedding
of the relevant low-energy fields.

Such a specification would amount to constructing a complete
microscopic dynamics from which the effective field content, spacetime
structure, and quantum description could all be recovered. In
principle, physical evolution could then be obtained directly by
iterating the fundamental deterministic map, with QFT retained as its
coarse-grained effective description. Determining $F$ is therefore not
a matter of fixing an additional effective parameter, but the more
ambitious problem of identifying the underlying dynamics of the full
theory. The model-independent results presented here constrain the
structure and infrared behavior of such a dynamics without
presupposing its explicit realization.

The running quantity $\hbar_{\rm eff}(\mu)$ should not be interpreted
as an independently varying fundamental constant. It parametrizes the
normalization of the effective canonical algebra at the
coarse-graining scale $\mu$ and measures the fraction of microscopic
information that remains unresolved at that scale. Its observable
content must therefore appear through scale-dependent
modifications of physical quantities, such as correlation functions,
propagators, spectra, or scattering amplitudes, rather than through an
isolated measurement of $\hbar_{\rm eff}$.

In the infrared, the cardinality of the unresolved equivalence classes
becomes large, $\hbar_{\rm eff}\to\hbar$, and the finite-difference
operators introduced here reduce to their continuum counterparts.
Standard unitary QFT is therefore recovered in the smooth low-energy
regime, with no unsuppressed modification of its predictions. Possible
deviations may instead arise when the observational resolution
approaches the fundamental scales associated with the microscopic
proper-time update or with the finite displacement in field-
configuration space.

For a specified microscopic realization, these deviations would
provide falsifiability criteria. Deformations of the canonical
algebra, residual nonunitary contributions, and ultraviolet
modifications would originate from the same microscopic dynamics and
coarse-graining prescription, and could not be chosen independently.
A realization would be excluded if it failed to recover ordinary
unitary and relativistic QFT in the infrared, or if its predicted
corrections were incompatible with observation. The possible
applications to loop amplitudes, horizon radiation, and ultraviolet
propagation mentioned above should therefore be understood as
directions for future quantitative investigation, rather than as
predictions already derived in the present model-independent analysis.

An explicit example of this phenomenological strategy is provided by
the binary-sequence construction of Ref.~\cite{Powers:2023ydg}. In
that approach, quantum probabilities are obtained by counting
underlying ontic configurations, leading to small finite-system
deviations from standard quantum predictions and to a proposed
optical test. Although its microscopic implementation differs from
the one considered here, it illustrates how a discrete ontological
model can become experimentally distinguishable once its microscopic
states and measurement prescription are specified.

A deterministic microscopic dynamics naturally invites comparison with Bell-type
and Kochen--Specker-type no-go theorems. In the present framework, however,
quantum observables emerge only after coarse-graining and are not identified
with pre-existing properties of the microscopic configurations. The framework
therefore does not assume that all quantum observables have simultaneous definite
values independent of the measurement context.

Moreover, the recovery of ordinary QFT in the infrared already
ensures the emergence of its quantum correlations and operational
no-signalling properties. Bell and Kochen--Specker results do not
therefore impose an additional independent dynamical condition on
$F$ beyond the recovery of the correct infrared theory. Rather, they
constrain the interpretation of the microscopic variables and exclude
their identification with local, noncontextual values of the emergent
quantum observables. At the fundamental level, the system, apparatus,
and measurement settings belong to the same deterministic microscopic
history; their effective statistical independence need not therefore
be imposed as a property of the underlying dynamics. The present
framework does not evade the no-go theorems, but places their
assumptions at the appropriate emergent level.

\section{Summary and Outlook}
\label{end}

In this work, we have extended the deterministic proper-time framework
introduced in Ref.~\cite{Maiezza:2026wke} in several directions, advancing
the construction of a realistic emergent QFT from a deterministic substrate.

First, we have clarified the relation between coarse-graining and the
renormalization-group structure of the emergent QFT. At resolutions
sufficiently larger than the fundamental proper-time interval, changes in
macroscopic resolution lead to the ordinary Callan--Symanzik equation. The
specific beta functions and anomalous dimensions depend on the interactions
of the emergent continuum theory, while the deterministic substrate provides
a microscopic interpretation of the change of scale as a change in the number
of elementary updates that remain unresolved. As the fundamental scale is
approached, the ordinary renormalization-group flow no longer provides a
complete description of the scale dependence, and the running of
$\hbar_{\rm eff}$ characterizes the crossover toward the underlying
deterministic dynamics.

Second, we have connected the deformed canonical commutation relations and
the running effective Planck constant to the algebra of finite translations
in field-configuration space. A local representation of the elementary
reversible update leads simultaneously to a deformed Leibniz rule and to an
operator-valued field-momentum commutator. On states that do not resolve the
finite field-space displacement, the ordinary functional derivative and the
standard canonical algebra are recovered. After restriction to a
coarse-grained equivalence class, the symmetric finite-translation operator
reproduces the same effective quantization factor previously obtained from
statistical microstate counting. Therefore, the finite-update construction
provides an algebraic realization of the running of $\hbar_{\rm eff}$ within
the discrete deterministic framework.

One central extension of the present work concerns fermionic degrees of
freedom. The microscopic variables whose infrared representation is
fermionic remain ordinary labels of the deterministic state space, while the
Grassmann structure is introduced only at the level of the effective graded
algebra. Within the minimal factorized representation adopted here, the
symmetric operator associated with the universal finite update dresses the
Berezin derivative, and the graded Leibniz rule yields the corresponding
operator-valued canonical anticommutator. Projecting onto the same
equivalence class and using the same coarse-graining map as in the bosonic
sector gives the same effective Planck constant in both canonical
representations.

The bosonic and fermionic algebras remain different, as required by their
respective gradings, but their normalization is common because both represent
the same microscopic update and the same unresolved multiplicity. Within the
common deterministic dynamics, coarse-graining map, and minimal factorized
fermionic representation considered here, this equality provides a
consistency test of the framework. It also supplies the fermionic canonical
sector required for extending the emergent QFT beyond the single-field
construction considered previously.

We have also provided an exploratory discussion of two possible
phenomenological consequences of the deterministic crossover. The first is a
relative suppression of radiative corrections when the characteristic
proper-time resolution of a high-energy process approaches
$\tau_{\min}$. The second is a possible suppression of Hawking radiation as
the effective quantization scale decreases. These considerations identify
qualitative signatures and limiting behavior rather than complete
predictions. A quantitative scattering analysis requires a
process-dependent relation between the hard scale, the proper-time
resolution, and the coarse-graining factor. Likewise, the black-hole
application requires the determination of the relevant coarse-graining scale
in the semiclassical geometry, together with a derivation of the modified
emission spectrum and the corresponding greybody and backreaction effects.

The present results strengthen the emergent QFT construction in three
related respects. They provide a microscopic interpretation of its
renormalization-group structure, an operatorial realization of the
coarse-grained quantization scale, and a common canonical normalization for
bosonic and fermionic degrees of freedom. In this sense, the present work
does not introduce an independent extension alongside the previous
framework, but develops algebraic and field-theoretic structures that were
left open in Ref.~\cite{Maiezza:2026wke}.

A possible next step would be to formulate interacting gauge theories
within the same framework, studying how gauge symmetries, constraints,
and interactions are represented by the universal deterministic map
and preserved under coarse-graining. This is an extraordinarily harder
problem than the kinematical construction developed here: it
ultimately requires a realistic $F$ that reproduces an interacting
quantum field theory, including Standard-Model-type gauge
interactions, in the appropriate coarse-grained limit -- in essence, a
complete and working deterministic microscopic theory, not only its
canonical algebra. The present work only establishes necessary
algebraic conditions that any such $F$ must satisfy after
coarse-graining, namely the common bosonic and fermionic normalization
derived above.

A separate outlook concerns the gravitational sector. In
Ref.~\cite{Maiezza:2026wke}, general relativity emerged in the infrared
through the coarse-grained geometric construction and the application of
Lovelock's theorem in four spacetime dimensions \cite{lovelock1972four}.
The extension from the single emergent field considered there to the
complete microscopic configuration $\Phi_{\rm tot}$ is consistent with the
common deterministic update used in the present work. The dimensionality
of spacetime, however, remains a phenomenological input of that gravitational
construction. It is an open question whether four dimensions can be selected
internally by the deterministic substrate or whether dimensionality should
remain an input of the effective theory. Resolving this question is not
required for the bosonic and fermionic canonical construction developed here,
but it remains a possible direction for understanding more fully the emergence of the
infrared geometric description.

\section*{Acknowledgments}

The author thanks his family for their support.

\bibliography{biblio}


\begin{thebibliography}{49}
\ifx \bisbn   \undefined \def \bisbn  #1{ISBN #1}\fi
\ifx \binits  \undefined \def \binits#1{#1}\fi
\ifx \bauthor  \undefined \def \bauthor#1{#1}\fi
\ifx \batitle  \undefined \def \batitle#1{#1}\fi
\ifx \bjtitle  \undefined \def \bjtitle#1{#1}\fi
\ifx \bvolume  \undefined \def \bvolume#1{\textbf{#1}}\fi
\ifx \byear  \undefined \def \byear#1{#1}\fi
\ifx \bissue  \undefined \def \bissue#1{#1}\fi
\ifx \bfpage  \undefined \def \bfpage#1{#1}\fi
\ifx \blpage  \undefined \def \blpage #1{#1}\fi
\ifx \burl  \undefined \def \burl#1{\textsf{#1}}\fi
\ifx \doiurl  \undefined \def \doiurl#1{\url{https://doi.org/#1}}\fi
\ifx \betal  \undefined \def \betal{\textit{et al.}}\fi
\ifx \binstitute  \undefined \def \binstitute#1{#1}\fi
\ifx \binstitutionaled  \undefined \def \binstitutionaled#1{#1}\fi
\ifx \bctitle  \undefined \def \bctitle#1{#1}\fi
\ifx \beditor  \undefined \def \beditor#1{#1}\fi
\ifx \bpublisher  \undefined \def \bpublisher#1{#1}\fi
\ifx \bbtitle  \undefined \def \bbtitle#1{#1}\fi
\ifx \bedition  \undefined \def \bedition#1{#1}\fi
\ifx \bseriesno  \undefined \def \bseriesno#1{#1}\fi
\ifx \blocation  \undefined \def \blocation#1{#1}\fi
\ifx \bsertitle  \undefined \def \bsertitle#1{#1}\fi
\ifx \bsnm \undefined \def \bsnm#1{#1}\fi
\ifx \bsuffix \undefined \def \bsuffix#1{#1}\fi
\ifx \bparticle \undefined \def \bparticle#1{#1}\fi
\ifx \barticle \undefined \def \barticle#1{#1}\fi
\bibcommenthead
\ifx \bconfdate \undefined \def \bconfdate #1{#1}\fi
\ifx \botherref \undefined \def \botherref #1{#1}\fi
\ifx \url \undefined \def \url#1{\textsf{#1}}\fi
\ifx \bchapter \undefined \def \bchapter#1{#1}\fi
\ifx \bbook \undefined \def \bbook#1{#1}\fi
\ifx \bcomment \undefined \def \bcomment#1{#1}\fi
\ifx \oauthor \undefined \def \oauthor#1{#1}\fi
\ifx \citeauthoryear \undefined \def \citeauthoryear#1{#1}\fi
\ifx \endbibitem  \undefined \def \endbibitem {}\fi
\ifx \bconflocation  \undefined \def \bconflocation#1{#1}\fi
\ifx \arxivurl  \undefined \def \arxivurl#1{\textsf{#1}}\fi
\csname PreBibitemsHook\endcsname

\bibitem[\protect\citeauthoryear{Mead}{1964}]{Mead:1964zz}
\begin{barticle}
\bauthor{\bsnm{Mead}, \binits{C.A.}}:
\batitle{{Possible Connection Between Gravitation and Fundamental Length}}.
\bjtitle{Phys. Rev.}
\bvolume{135},
\bfpage{849}--\blpage{862}
(\byear{1964})
\doiurl{10.1103/PhysRev.135.B849}
\end{barticle}
\endbibitem

\bibitem[\protect\citeauthoryear{GARAY}{1995}]{GARAY_1995}
\begin{barticle}
\bauthor{\bsnm{GARAY}, \binits{L.J.}}:
\batitle{Quantum gravity and minimum length}.
\bjtitle{International Journal of Modern Physics A}
\bvolume{10}(\bissue{02}),
\bfpage{145}--\blpage{165}
(\byear{1995})
\doiurl{10.1142/s0217751x95000085}
\end{barticle}
\endbibitem

\bibitem[\protect\citeauthoryear{Kempf et~al.}{1995}]{Kempf_1995}
\begin{barticle}
\bauthor{\bsnm{Kempf}, \binits{A.}},
\bauthor{\bsnm{Mangano}, \binits{G.}},
\bauthor{\bsnm{Mann}, \binits{R.B.}}:
\batitle{Hilbert space representation of the minimal length uncertainty
  relation}.
\bjtitle{Physical Review D}
\bvolume{52}(\bissue{2}),
\bfpage{1108}--\blpage{1118}
(\byear{1995})
\doiurl{10.1103/physrevd.52.1108}
\end{barticle}
\endbibitem

\bibitem[\protect\citeauthoryear{Padmanabhan}{1997}]{Padmanabhan:1996ap}
\begin{barticle}
\bauthor{\bsnm{Padmanabhan}, \binits{T.}}:
\batitle{{Duality and zero point length of space-time}}.
\bjtitle{Phys. Rev. Lett.}
\bvolume{78},
\bfpage{1854}--\blpage{1857}
(\byear{1997})
\doiurl{10.1103/PhysRevLett.78.1854}
{\href{https://arxiv.org/abs/hep-th/9608182}{{arXiv:hep-th/9608182}}}
\end{barticle}
\endbibitem

\bibitem[\protect\citeauthoryear{Modesto}{2009}]{Modesto_2009}
\begin{barticle}
\bauthor{\bsnm{Modesto}, \binits{L.}}:
\batitle{Fractal spacetime from the area spectrum}.
\bjtitle{Classical and Quantum Gravity}
\bvolume{26}(\bissue{24}),
\bfpage{242002}
(\byear{2009})
\doiurl{10.1088/0264-9381/26/24/242002}
\end{barticle}
\endbibitem

\bibitem[\protect\citeauthoryear{Nicolini and Niedner}{2011}]{Nicolini_2011}
\begin{botherref}
\oauthor{\bsnm{Nicolini}, \binits{P.}},
\oauthor{\bsnm{Niedner}, \binits{B.}}:
Hausdorff dimension of a particle path in a quantum manifold.
Physical Review D
\textbf{83}(2)
(2011)
\doiurl{10.1103/physrevd.83.024017}
\end{botherref}
\endbibitem

\bibitem[\protect\citeauthoryear{Bosso et~al.}{2023}]{Bosso_2023}
\begin{botherref}
\oauthor{\bsnm{Bosso}, \binits{P.}},
\oauthor{\bsnm{Petruzziello}, \binits{L.}},
\oauthor{\bsnm{Wagner}, \binits{F.}}:
Minimal length: A cut-off in disguise?
Physical Review D
\textbf{107}(12)
(2023)
\doiurl{10.1103/physrevd.107.126009}
\end{botherref}
\endbibitem

\bibitem[\protect\citeauthoryear{Bosso}{2024}]{Bosso_2024}
\begin{botherref}
\oauthor{\bsnm{Bosso}, \binits{P.}}:
Minimal-length quantum field theory: a first-principle approach.
The European Physical Journal C
\textbf{84}(9)
(2024)
\doiurl{10.1140/epjc/s10052-024-13281-9}
\end{botherref}
\endbibitem

\bibitem[\protect\citeauthoryear{D'Agostino et~al.}{2026}]{DAgostino:2025axy}
\begin{barticle}
\bauthor{\bsnm{D'Agostino}, \binits{R.}},
\bauthor{\bsnm{Bosso}, \binits{P.}},
\bauthor{\bsnm{Luciano}, \binits{G.G.}}:
\batitle{{From minimal-length quantum theory to modified gravity}}.
\bjtitle{Phys. Lett. B}
\bvolume{878},
\bfpage{140519}
(\byear{2026})
\doiurl{10.1016/j.physletb.2026.140519}
{\href{https://arxiv.org/abs/2511.14869}{{arXiv:2511.14869}}}
{[gr-qc]}
\end{barticle}
\endbibitem

\bibitem[\protect\citeauthoryear{Maiezza and Vasquez}{2026}]{Maiezza:2026wrp}
\begin{barticle}
\bauthor{\bsnm{Maiezza}, \binits{A.}},
\bauthor{\bsnm{Vasquez}, \binits{J.C.}}:
\batitle{{Minimal proper-time in quantum field theory}}.
\bjtitle{Nucl. Phys. B}
\bvolume{1025},
\bfpage{117409}
(\byear{2026})
\doiurl{10.1016/j.nuclphysb.2026.117409}
{\href{https://arxiv.org/abs/2602.00045}{{arXiv:2602.00045}}}
{[hep-th]}
\end{barticle}
\endbibitem

\bibitem[\protect\citeauthoryear{Maggiore}{1993}]{Maggiore:1993rv}
\begin{barticle}
\bauthor{\bsnm{Maggiore}, \binits{M.}}:
\batitle{{A Generalized uncertainty principle in quantum gravity}}.
\bjtitle{Phys. Lett. B}
\bvolume{304},
\bfpage{65}--\blpage{69}
(\byear{1993})
\doiurl{10.1016/0370-2693(93)91401-8}
{\href{https://arxiv.org/abs/hep-th/9301067}{{arXiv:hep-th/9301067}}}
\end{barticle}
\endbibitem

\bibitem[\protect\citeauthoryear{Adler and Santiago}{1999}]{Adler:1999bu}
\begin{barticle}
\bauthor{\bsnm{Adler}, \binits{R.J.}},
\bauthor{\bsnm{Santiago}, \binits{D.I.}}:
\batitle{{On gravity and the uncertainty principle}}.
\bjtitle{Mod. Phys. Lett. A}
\bvolume{14},
\bfpage{1371}
(\byear{1999})
\doiurl{10.1142/S0217732399001462}
{\href{https://arxiv.org/abs/gr-qc/9904026}{{arXiv:gr-qc/9904026}}}
\end{barticle}
\endbibitem

\bibitem[\protect\citeauthoryear{Ong}{2018}]{Ong:2018zqn}
\begin{barticle}
\bauthor{\bsnm{Ong}, \binits{Y.C.}}:
\batitle{{Generalized Uncertainty Principle, Black Holes, and White Dwarfs: A
  Tale of Two Infinities}}.
\bjtitle{JCAP}
\bvolume{09},
\bfpage{015}
(\byear{2018})
\doiurl{10.1088/1475-7516/2018/09/015}
{\href{https://arxiv.org/abs/1804.05176}{{arXiv:1804.05176}}}
{[gr-qc]}
\end{barticle}
\endbibitem

\bibitem[\protect\citeauthoryear{Ong}{2023}]{Ong:2023jkp}
\begin{barticle}
\bauthor{\bsnm{Ong}, \binits{Y.C.}}:
\batitle{{A critique on some aspects of GUP effective metric}}.
\bjtitle{Eur. Phys. J. C}
\bvolume{83}(\bissue{3}),
\bfpage{209}
(\byear{2023})
\doiurl{10.1140/epjc/s10052-023-11360-x}
{\href{https://arxiv.org/abs/2303.10719}{{arXiv:2303.10719}}}
{[gr-qc]}
\end{barticle}
\endbibitem

\bibitem[\protect\citeauthoryear{Bosso et~al.}{2023}]{Bosso:2023aht}
\begin{barticle}
\bauthor{\bsnm{Bosso}, \binits{P.}},
\bauthor{\bsnm{Luciano}, \binits{G.G.}},
\bauthor{\bsnm{Petruzziello}, \binits{L.}},
\bauthor{\bsnm{Wagner}, \binits{F.}}:
\batitle{{30 years in: Quo vadis generalized uncertainty principle?}}
\bjtitle{Class. Quant. Grav.}
\bvolume{40}(\bissue{19}),
\bfpage{195014}
(\byear{2023})
\doiurl{10.1088/1361-6382/acf021}
{\href{https://arxiv.org/abs/2305.16193}{{arXiv:2305.16193}}}
{[gr-qc]}
\end{barticle}
\endbibitem

\bibitem[\protect\citeauthoryear{Ong}{2025}]{Ong:2025ent}
\begin{barticle}
\bauthor{\bsnm{Ong}, \binits{Y.C.}}:
\batitle{{GUP Effective metric without GUP: Implications for the sign of GUP
  parameter and quantum bounce}}.
\bjtitle{Phys. Lett. B}
\bvolume{870},
\bfpage{139936}
(\byear{2025})
\doiurl{10.1016/j.physletb.2025.139936}
{\href{https://arxiv.org/abs/2505.07972}{{arXiv:2505.07972}}}
{[gr-qc]}
\end{barticle}
\endbibitem

\bibitem[\protect\citeauthoryear{Adler}{2004}]{adler2004quantum}
\begin{bbook}
\bauthor{\bsnm{Adler}, \binits{S.L.}}:
\bbtitle{Quantum Theory as an Emergent Phenomenon: The Statistical Mechanics of
  Matrix Models as the Precursor of Quantum Field Theory}.
\bpublisher{Cambridge University Press}, \blocation{???}
(\byear{2004})
\end{bbook}
\endbibitem

\bibitem[\protect\citeauthoryear{Volovik}{2010}]{Volovik:2009xs}
\begin{barticle}
\bauthor{\bsnm{Volovik}, \binits{G.E.}}:
\batitle{{h-bar as parameter of Minkowski metric in effective theory}}.
\bjtitle{JETP Lett.}
\bvolume{90},
\bfpage{697}--\blpage{704}
(\byear{2010})
\doiurl{10.1134/S0021364009230027}
{\href{https://arxiv.org/abs/0904.1965}{{arXiv:0904.1965}}}
{[gr-qc]}
\end{barticle}
\endbibitem

\bibitem[\protect\citeauthoryear{Hossenfelder}{2013}]{Hossenfelder:2012uy}
\begin{barticle}
\bauthor{\bsnm{Hossenfelder}, \binits{S.}}:
\batitle{{A possibility to solve the problems with quantizing gravity}}.
\bjtitle{Phys. Lett. B}
\bvolume{725},
\bfpage{473}--\blpage{476}
(\byear{2013})
\doiurl{10.1016/j.physletb.2013.07.037}
{\href{https://arxiv.org/abs/1208.5874}{{arXiv:1208.5874}}}
{[gr-qc]}
\end{barticle}
\endbibitem

\bibitem[\protect\citeauthoryear{'t~Hooft}{2003}]{tHooft:2001qty}
\begin{barticle}
\bauthor{\bsnm{Hooft}, \binits{G.}}:
\batitle{{Determinism in free bosons}}.
\bjtitle{Int. J. Theor. Phys.}
\bvolume{42},
\bfpage{355}--\blpage{361}
(\byear{2003})
\doiurl{10.1023/A:1024459703072}
{\href{https://arxiv.org/abs/hep-th/0104080}{{arXiv:hep-th/0104080}}}
\end{barticle}
\endbibitem

\bibitem[\protect\citeauthoryear{Palmer}{2009}]{Palmer:2008jh}
\begin{barticle}
\bauthor{\bsnm{Palmer}, \binits{T.N.}}:
\batitle{{The Invariant Set Hypothesis: A New Geometric Framework for the
  Foundations of Quantum Theory and the Role Played by Gravity}}.
\bjtitle{Proc. Roy. Soc. Lond. A}
\bvolume{465},
\bfpage{3187}--\blpage{3207}
(\byear{2009})
\doiurl{10.1098/rspa.2009.0080}
{\href{https://arxiv.org/abs/0812.1148}{{arXiv:0812.1148}}}
{[quant-ph]}
\end{barticle}
\endbibitem

\bibitem[\protect\citeauthoryear{Hall}{2010}]{Hall:2010zzf}
\begin{barticle}
\bauthor{\bsnm{Hall}, \binits{M.J.W.}}:
\batitle{{Local deterministic model of singlet state correlations based on
  relaxing measurement independence}}.
\bjtitle{Phys. Rev. Lett.}
\bvolume{105},
\bfpage{250404}
(\byear{2010})
\doiurl{10.1103/PhysRevLett.105.250404}
{\href{https://arxiv.org/abs/1007.5518}{{arXiv:1007.5518}}}
{[quant-ph]}.
\bcomment{[Erratum: Phys.Rev.Lett. 116, 219902 (2016)]}
\end{barticle}
\endbibitem

\bibitem[\protect\citeauthoryear{Hossenfelder and
  Palmer}{2020}]{Hossenfelder:2019shy}
\begin{barticle}
\bauthor{\bsnm{Hossenfelder}, \binits{S.}},
\bauthor{\bsnm{Palmer}, \binits{T.N.}}:
\batitle{{Rethinking Superdeterminism}}.
\bjtitle{Front. in Phys.}
\bvolume{8},
\bfpage{139}
(\byear{2020})
\doiurl{10.3389/fphy.2020.00139}
{\href{https://arxiv.org/abs/1912.06462}{{arXiv:1912.06462}}}
{[quant-ph]}
\end{barticle}
\endbibitem

\bibitem[\protect\citeauthoryear{Donadi and
  Hossenfelder}{2022}]{Donadi:2020aqz}
\begin{barticle}
\bauthor{\bsnm{Donadi}, \binits{S.}},
\bauthor{\bsnm{Hossenfelder}, \binits{S.}}:
\batitle{{Toy model for local and deterministic wave-function collapse}}.
\bjtitle{Phys. Rev. A}
\bvolume{106}(\bissue{2}),
\bfpage{022212}--\blpage{102221214}
(\byear{2022})
\doiurl{10.1103/PhysRevA.106.022212}
{\href{https://arxiv.org/abs/2010.01327}{{arXiv:2010.01327}}}
{[quant-ph]}
\end{barticle}
\endbibitem

\bibitem[\protect\citeauthoryear{'t~Hooft}{2021}]{tHooft:2020qfg}
\begin{barticle}
\bauthor{\bsnm{Hooft}, \binits{G.}}:
\batitle{{Fast Vacuum Fluctuations and the Emergence of Quantum Mechanics}}.
\bjtitle{Found. Phys.}
\bvolume{51}(\bissue{3}),
\bfpage{63}
(\byear{2021})
\doiurl{10.1007/s10701-021-00464-7}
{\href{https://arxiv.org/abs/2010.02019}{{arXiv:2010.02019}}}
{[quant-ph]}
\end{barticle}
\endbibitem

\bibitem[\protect\citeauthoryear{'t~Hooft}{2020}]{tHooft:2020tuu}
\begin{barticle}
\bauthor{\bsnm{Hooft}, \binits{G.}}:
\batitle{{Deterministic Quantum Mechanics: The Mathematical Equations}}.
\bjtitle{Front. in Phys.}
\bvolume{8},
\bfpage{253}
(\byear{2020})
\doiurl{10.3389/fphy.2020.00253}
{\href{https://arxiv.org/abs/2005.06374}{{arXiv:2005.06374}}}
{[quant-ph]}
\end{barticle}
\endbibitem

\bibitem[\protect\citeauthoryear{Powers and Stojkovic}{2022}]{Powers:2021rfg}
\begin{barticle}
\bauthor{\bsnm{Powers}, \binits{S.}},
\bauthor{\bsnm{Stojkovic}, \binits{D.}}:
\batitle{{An alternative formalism for modeling spin}}.
\bjtitle{Eur. Phys. J. C}
\bvolume{82},
\bfpage{690}
(\byear{2022})
\doiurl{10.1140/epjc/s10052-022-10652-y}
{\href{https://arxiv.org/abs/2110.13617}{{arXiv:2110.13617}}}
{[physics.gen-ph]}
\end{barticle}
\endbibitem

\bibitem[\protect\citeauthoryear{Hance and Hossenfelder}{2022}]{Hance:2022juc}
\begin{barticle}
\bauthor{\bsnm{Hance}, \binits{J.R.}},
\bauthor{\bsnm{Hossenfelder}, \binits{S.}}:
\batitle{{Bell's theorem allows local theories of quantum mechanics}}.
\bjtitle{Nature Phys.}
\bvolume{18},
\bfpage{1382}
(\byear{2022})
\doiurl{10.1038/s41567-022-01831-5}
{\href{https://arxiv.org/abs/2211.01331}{{arXiv:2211.01331}}}
{[quant-ph]}
\end{barticle}
\endbibitem

\bibitem[\protect\citeauthoryear{Palmer}{2024}]{Palmer:2023vfw}
\begin{barticle}
\bauthor{\bsnm{Palmer}, \binits{T.}}:
\batitle{{Superdeterminism without Conspiracy {\textdagger}}}.
\bjtitle{Universe}
\bvolume{10}(\bissue{1}),
\bfpage{47}
(\byear{2024})
\doiurl{10.3390/universe10010047}
{\href{https://arxiv.org/abs/2308.11262}{{arXiv:2308.11262}}}
{[quant-ph]}
\end{barticle}
\endbibitem

\bibitem[\protect\citeauthoryear{Arroyo}{2025}]{Arroyo:2024saq}
\begin{barticle}
\bauthor{\bsnm{Arroyo}, \binits{E.A.}}:
\batitle{{A family of deterministic models for singlet quantum state
  correlations}}.
\bjtitle{J. Phys. A}
\bvolume{58}(\bissue{24}),
\bfpage{245301}
(\byear{2025})
\doiurl{10.1088/1751-8121/ade104}
{\href{https://arxiv.org/abs/2408.09579}{{arXiv:2408.09579}}}
{[quant-ph]}
\end{barticle}
\endbibitem

\bibitem[\protect\citeauthoryear{Maiezza}{2026}]{Maiezza:2026wke}
\begin{barticle}
\bauthor{\bsnm{Maiezza}, \binits{A.}}:
\batitle{{Minimal proper time and deterministic microstates: emergent quantum
  fields and relativistic spacetime}}.
\bjtitle{Eur. Phys. J. C}
\bvolume{86}(\bissue{7}),
\bfpage{829}
(\byear{2026})
\doiurl{10.1140/epjc/s10052-026-16089-x}
{\href{https://arxiv.org/abs/2607.03605}{{arXiv:2607.03605}}}
{[hep-th]}
\end{barticle}
\endbibitem

\bibitem[\protect\citeauthoryear{Petruzziello and
  Illuminati}{2021}]{Petruzziello:2020wkd}
\begin{barticle}
\bauthor{\bsnm{Petruzziello}, \binits{L.}},
\bauthor{\bsnm{Illuminati}, \binits{F.}}:
\batitle{{Quantum gravitational decoherence from fluctuating minimal length and
  deformation parameter at the Planck scale}}.
\bjtitle{Nature Commun.}
\bvolume{12}(\bissue{1}),
\bfpage{4449}
(\byear{2021})
\doiurl{10.1038/s41467-021-24711-7}
{\href{https://arxiv.org/abs/2011.01255}{{arXiv:2011.01255}}}
{[gr-qc]}
\end{barticle}
\endbibitem

\bibitem[\protect\citeauthoryear{Bombelli et~al.}{1987}]{Bombelli:1987aa}
\begin{barticle}
\bauthor{\bsnm{Bombelli}, \binits{L.}},
\bauthor{\bsnm{Lee}, \binits{J.}},
\bauthor{\bsnm{Meyer}, \binits{D.}},
\bauthor{\bsnm{Sorkin}, \binits{R.}}:
\batitle{{Space-Time as a Causal Set}}.
\bjtitle{Phys. Rev. Lett.}
\bvolume{59},
\bfpage{521}--\blpage{524}
(\byear{1987})
\doiurl{10.1103/PhysRevLett.59.521}
\end{barticle}
\endbibitem

\bibitem[\protect\citeauthoryear{Johnston}{2010}]{Johnston:2010su}
\begin{botherref}
\oauthor{\bsnm{Johnston}, \binits{S.P.}}:
{Quantum Fields on Causal Sets}.
Other thesis
(October 2010)
\end{botherref}
\endbibitem

\bibitem[\protect\citeauthoryear{Powers and
  Stojkovic}{2026}]{powers2026eventcentricapproachmodeling}
\begin{botherref}
\oauthor{\bsnm{Powers}, \binits{S.}},
\oauthor{\bsnm{Stojkovic}, \binits{D.}}:
An event centric approach to modeling quantum systems
(2026).
\url{https://arxiv.org/abs/2306.14922}
\end{botherref}
\endbibitem

\bibitem[\protect\citeauthoryear{Volovik}{2026}]{volovik2026quantumclassicalmechanicsvs}
\begin{botherref}
\oauthor{\bsnm{Volovik}, \binits{G.E.}}:
Quantum and Classical mechanics vs QFT
(2026).
\url{https://arxiv.org/abs/2602.21243}
\end{botherref}
\endbibitem

\bibitem[\protect\citeauthoryear{Bakr and Zhang}{2026}]{Bakr:2026ngx}
\begin{botherref}
\oauthor{\bsnm{Bakr}, \binits{M.}},
\oauthor{\bsnm{Zhang}, \binits{T.}}:
{Scalar Finite-Proper-Time Field Theory as a Complete-History Spectral
  Calculus}
(2026)
{\href{https://arxiv.org/abs/2608.11721}{{arXiv:2608.11721}}}
{[hep-th]}
\end{botherref}
\endbibitem

\bibitem[\protect\citeauthoryear{Magueijo and
  Smolin}{2003}]{MagueijoSmolin2003}
\begin{barticle}
\bauthor{\bsnm{Magueijo}, \binits{J.}},
\bauthor{\bsnm{Smolin}, \binits{L.}}:
\batitle{Generalized lorentz invariance with an invariant energy scale}.
\bjtitle{Physical Review D}
\bvolume{67},
\bfpage{044004}
(\byear{2003})
{\href{https://arxiv.org/abs/gr-qc/0207085}{{gr-qc/0207085}}}
\end{barticle}
\endbibitem

\bibitem[\protect\citeauthoryear{Volovik}{2003}]{Volovik2003}
\begin{bbook}
\bauthor{\bsnm{Volovik}, \binits{G.E.}}:
\bbtitle{The Universe in a Helium Droplet}.
\bpublisher{Oxford University Press}, \blocation{???}
(\byear{2003})
\end{bbook}
\endbibitem

\bibitem[\protect\citeauthoryear{'t~Hooft}{2016}]{thooft2016cellular}
\begin{bbook}
\bauthor{\bsnm{Hooft}, \binits{G.}}:
\bbtitle{The Cellular Automaton Interpretation of Quantum Mechanics}.
\bsertitle{Fundamental Theories of Physics},
vol. \bseriesno{185}.
\bpublisher{Springer},
\blocation{Cham, Switzerland}
(\byear{2016}).
\doiurl{10.1007/978-3-319-41285-6}
\end{bbook}
\endbibitem

\bibitem[\protect\citeauthoryear{Hawking}{1974}]{Hawking:1974rv}
\begin{barticle}
\bauthor{\bsnm{Hawking}, \binits{S.W.}}:
\batitle{{Black hole explosions}}.
\bjtitle{Nature}
\bvolume{248},
\bfpage{30}--\blpage{31}
(\byear{1974})
\doiurl{10.1038/248030a0}
\end{barticle}
\endbibitem

\bibitem[\protect\citeauthoryear{Hawking}{1977}]{Hawking:1977wz}
\begin{barticle}
\bauthor{\bsnm{Hawking}, \binits{S.W.}}:
\batitle{{The Quantum Mechanics of Black Holes}}.
\bjtitle{Sci. Am.}
\bvolume{236},
\bfpage{34}--\blpage{49}
(\byear{1977})
\doiurl{10.1038/scientificamerican0177-34}
\end{barticle}
\endbibitem

\bibitem[\protect\citeauthoryear{Parikh and Wilczek}{2000}]{Parikh:1999mf}
\begin{barticle}
\bauthor{\bsnm{Parikh}, \binits{M.K.}},
\bauthor{\bsnm{Wilczek}, \binits{F.}}:
\batitle{{Hawking radiation as tunneling}}.
\bjtitle{Phys. Rev. Lett.}
\bvolume{85},
\bfpage{5042}--\blpage{5045}
(\byear{2000})
\doiurl{10.1103/PhysRevLett.85.5042}
{\href{https://arxiv.org/abs/hep-th/9907001}{{arXiv:hep-th/9907001}}}
\end{barticle}
\endbibitem

\bibitem[\protect\citeauthoryear{Hossenfelder et~al.}{2003}]{Hossenfelder2003}
\begin{barticle}
\bauthor{\bsnm{Hossenfelder}, \binits{S.}},
\bauthor{\bsnm{Bleicher}, \binits{M.}},
\bauthor{\bsnm{Hofmann}, \binits{S.}},
\bauthor{\bsnm{Ruppert}, \binits{J.}},
\bauthor{\bsnm{Scherer}, \binits{S.}},
\bauthor{\bsnm{St{\"o}cker}, \binits{H.}}:
\batitle{Signatures in the planck regime}.
\bjtitle{Physics Letters B}
\bvolume{575},
\bfpage{85}--\blpage{99}
(\byear{2003})
{\href{https://arxiv.org/abs/hep-th/0305262}{{hep-th/0305262}}}
\end{barticle}
\endbibitem

\bibitem[\protect\citeauthoryear{Hossenfelder}{2004}]{Hossenfelder2004}
\begin{barticle}
\bauthor{\bsnm{Hossenfelder}, \binits{S.}}:
\batitle{Interpretation of remnant energy in black hole evaporation}.
\bjtitle{Physical Review D}
\bvolume{70},
\bfpage{104003}
(\byear{2004})
{\href{https://arxiv.org/abs/hep-th/0405196}{{hep-th/0405196}}}
\end{barticle}
\endbibitem

\bibitem[\protect\citeauthoryear{Scardigli}{1999}]{Scardigli1999}
\begin{barticle}
\bauthor{\bsnm{Scardigli}, \binits{F.}}:
\batitle{Generalized uncertainty principle in quantum gravity from micro-black
  hole gedanken experiment}.
\bjtitle{Physics Letters B}
\bvolume{452}(\bissue{1-2}),
\bfpage{39}--\blpage{44}
(\byear{1999})
{\href{https://arxiv.org/abs/hep-th/9904025}{{hep-th/9904025}}}
\end{barticle}
\endbibitem

\bibitem[\protect\citeauthoryear{Adler et~al.}{2001}]{Adler2001}
\begin{barticle}
\bauthor{\bsnm{Adler}, \binits{R.J.}},
\bauthor{\bsnm{Chen}, \binits{P.}},
\bauthor{\bsnm{Santiago}, \binits{D.I.}}:
\batitle{The generalized uncertainty principle and black hole remnants}.
\bjtitle{General Relativity and Gravitation}
\bvolume{33},
\bfpage{2101}--\blpage{2108}
(\byear{2001})
{\href{https://arxiv.org/abs/gr-qc/0106080}{{gr-qc/0106080}}}
\end{barticle}
\endbibitem

\bibitem[\protect\citeauthoryear{Powers et~al.}{2025}]{Powers:2023ydg}
\begin{barticle}
\bauthor{\bsnm{Powers}, \binits{S.}},
\bauthor{\bsnm{Xu}, \binits{G.}},
\bauthor{\bsnm{Fotso}, \binits{H.}},
\bauthor{\bsnm{Thomay}, \binits{T.}},
\bauthor{\bsnm{Stojkovic}, \binits{D.}}:
\batitle{{Statistical model for quantum spin and photon number states}}.
\bjtitle{Phys. Rev. A}
\bvolume{111}(\bissue{1}),
\bfpage{012217}
(\byear{2025})
\doiurl{10.1103/PhysRevA.111.012217}
{\href{https://arxiv.org/abs/2304.13535}{{arXiv:2304.13535}}}
{[quant-ph]}
\end{barticle}
\endbibitem

\bibitem[\protect\citeauthoryear{Lovelock}{1972}]{lovelock1972four}
\begin{barticle}
\bauthor{\bsnm{Lovelock}, \binits{D.}}:
\batitle{The four-dimensionality of space and the einstein tensor}.
\bjtitle{Journal of Mathematical Physics}
\bvolume{13}(\bissue{6}),
\bfpage{874}--\blpage{876}
(\byear{1972})
\end{barticle}
\endbibitem

\end{thebibliography}

\end{document}